%% file: ms5.tex
\documentclass[iop]{emulateapj}

\bibstyle{ApJ}
\usepackage{epsfig}
\usepackage{amsmath}
\usepackage{natbib}
\usepackage{graphicx}

\newcommand\Msun  {${\rm M}_\odot$}
\newcommand\msun{\rm M_{\odot}}
\newcommand\mdot  {\dot{M}}
\newcommand\eg    {{\it e.g.\ }}

\newcommand\etal  {{\it et al.\ }}
\newcommand\kms{\rm \, km s^{-1}}

\begin{document}

\title{Competitive Accretion in Sheet Geometry and
the Stellar IMF}
\author{Wen-Hsin Hsu\altaffilmark{1}, Lee Hartmann\altaffilmark{1}, 
Fabian Heitsch\altaffilmark{2} and Gilberto C. G{\'o}mez\altaffilmark{3}}
\altaffiltext{1}{Dept. of Astronomy, University of Michigan, 500
Church St., Ann Arbor, MI 48109}
\altaffiltext{2}{Dept. of Physics \& Astronomy, University of North Carolina
at Chapel Hill, Chapel Hill, NC 27599-3255}
\altaffiltext{3}{Centro de Radioastronom\'{i}a y Astrof\'{i}sica, Universidad Nacional Aut\'onoma de M\'exico, Apdo. Postal 72-3
(Xangari), Morelia, Michoac\'{a}n 58089, Mexico}
\lefthead{Hsu et al.}
\shorttitle{Competitive Accretion in Sheet Geometry and the IMF}
\shortauthors{Hsu et al.}

\begin{abstract}
We report a set of numerical experiments aimed at 
addressing the applicability of competitive accretion 
to explain the high-mass end of the stellar initial mass function
in a sheet geometry with shallow gravitational potential, 
in contrast to most previous simulations which have assumed formation
in a cluster gravitational potential.
Our flat cloud geometry is motivated by models of molecular cloud formation
due to large-scale flows in the interstellar medium.
The experiments consisted of SPH simulations of gas accretion onto 
sink particles formed rapidly from Jeans-unstable
dense clumps placed randomly in
the finite sheet.  These simplifications allow us to study
accretion with a minimum of free parameters, and to 
develop better statistics on the resulting mass spectra.  
We considered both clumps of
equal mass and gaussian distributions of masses,
and either uniform or spatially-varying gas densities.
In all cases, the sink mass function develops a power law 
tail at high masses, with $dN/dlog M \propto M^{-\Gamma}$. 
The accretion rates of individual sinks follow $\dot{M} 
\propto M^2$ at high masses; this results in a continual
flattening of the slope of the mass function 
towards an asymptotic form $\Gamma \sim 1$
(where the Salpeter slope is $\Gamma = 1.35$). 
The asymptotic limit is most rapidly reached when starting from
a relatively broad distribution of initial sink masses.  In general
the resulting upper mass slope is correlated with the maximum
sink mass; higher sink masses are found in simulations with
flatter upper mass slopes.  Although these simulations are 
of a highly idealized situation, the results suggest that
competitive accretion may be relevant in a wider variety 
of environments than previously considered, and in particular
that the upper mass distribution may generally evolve towards
a limiting value of $\Gamma \sim 1$.

\end{abstract}

\section{Introduction}

The stellar initial mass function (IMF) among other things determines
the fraction of stellar populations in massive stars; 
this in turn affects the production of heavy elements, 
the stellar feedback of energy into the 
ISM, and the evolution of galaxies.
\citet{Salpeter55} first pointed out the power law distribution in 
the ``original mass function"; subsequent observational 
work has established the general form 
of the IMF, which at high masses is still comparable to 
the ``Salpeter slope" $\Gamma$, where $dN/d\log M = 
M^{-\Gamma}$, $\Gamma = 1.35$. The most 
widely used functional form is a power-law distribution or 
a combination of power-law distribution at different mass 
ranges. Other widely used forms of the 
IMF include log-normal distributions and combination of 
power-law and log-normal distribution 
\citep[e.g.][]{Chabrier03, BCM2010}. As \citet{BLZ07} noted, the 
essential features of the IMF include a peak at a mass of 
a few tenths of \Msun \ and a declining power-law tail 
toward higher masses. 

While the origin of the IMF remains a matter of extensive debate,
two general ideas have come to prominence in 
recent years (e.g., \citealt{Clarke09}).  The first supposes
that the mass spectrum of dense structures within star-forming clouds,
suggested to be the result of supersonic turbulence,
more or less directly maps into the stellar mass distribution (e.g., \citealt{Padoan02, KleinPPV}).  In these models
the IMF results from local mass reservoirs that are relatively
isolated \citep{Padoan07, Hennebelle08},
possibly affected by gravity \citep{Klessen00, Klessen01}.  
The second type of model
invokes two processes to produce the IMF; the low-mass end
is determined by turbulence and thermal physics, qualitatively
similar to the first picture, but the high-mass ``tail'' is a result of
continuing accretion from a mass reservoir
(e.g., \citealt{Zinnecker82, BBC01,BCB01, BLZ07}).
Thus the accumulation of material by the most massive stars
is the result of non-isolated accretion, from size scales
greater than the local Jeans length. 
The process resulting in producing
the high-mass end of the IMF in this approach is
usually called ``competitive accretion'' (CA).

As summarized by \citet{Clark09} and \citet{BLZ07}, 
the high-mass power-law tail in CA simulations typically arises from
formation in a stellar cluster; the potential well 
results in high gas densities near the center, helping
to feed material into the most massive objects
(see also \citealt{Bate09}).
Bonnell \etal (2001b) \nocite{BCB01} found that the slope of the mass function
depended upon whether the gravitational potential was dominated
by gas - in which case they found an asymptotic limit of
$\Gamma = 0.5$, due to tidal lobe limitation of mass accretion;
or by stars, in which case the asymptotic limit
was $\Gamma = 1$, where Bondi-Hoyle accretion 
dominates.  The latter is consistent with the analysis of Zinnecker
(1982), who showed that $\Gamma = 1$ results asympotically from
accretion rates which scale as $\mdot \propto M^2$.

These investigations suggest that CA can account for the high-mass end of
the IMF in clusters.  However, while most stars form in clusters, 
a non-negligible number do not, at least in the solar neighborhood. 
In addition, the properties of clusters vary widely, with most being
relatively small (Lada \& Lada 2003\nocite{Lada03}); this raises the question
as to whether the IMF might be affected by the mass of the cluster.
Moreover, the initial states and evolution of protocluster clouds 
and clusters are uncertain; current
assumptions range from relatively slow evolution in a roughly virialized
condition (e.g., Tan, Krumholz \& McKee 2006\nocite{Tan06}) to the
the opposite assumption of rapid gravitational collapse
(e.g., Tobin \etal 2009\nocite{Tobin09}; Proszkow \etal 2009\nocite{Proszkow09}).
We are therefore motivated to investigate a schematic model of competitive
accretion which does not employ the assumption of formation in an
initially clustered environment.
In addition, we wish to adopt a simple initial physical model with
as few parameters as possible to isolate the most important properties for
producing the high-mass IMF.

In this paper we report a set of numerical simulations
in a simplified model to address some general aspects of competitive
accretion.  Our results suggest that values of $\Gamma$ close to 
the Salpeter slope
can result in a wider variety of environments than previously
discussed; they also suggest that the value of $\Gamma$ may 
be correlated with
the maximum mass achieved through CA.
These findings suggest additional new approaches for numerical
simulations of the production of stellar IMFs.

\section{Model and Methods}

Our initial setup is motivated by our models of
molecular cloud formation as a result of large-scale flows in the interstellar
medium \citep{Heitsch06,Vazquez06,Heitsch08a,Heitsch08c,Heitsch08b}. 
In these models the dense material formed in
post-shock gas is geometrically thin rather than spherical, due to
post-shock compression by large-scale flows (e.g., Hartmann, Ballesteros-Paredes 
\& Bergin 2001\nocite{Hartmann01}).  
As there is no particular mechanism which would enforce virialization,
the cloud as a whole collapses laterally under gravity; eventually, much
if not most of the supersonic motion in the cloud is due to acceleration by
the cloud's self-gravity, rather than the initial turbulent velocities
injected during cloud formation (e.g., Heitsch \etal 2008; Heitsch \& Hartmann 2008).
The most important role of 
this mostly gravitationally driven 
turbulence in the post-shock gas is to provide density enhancements which
can gravitationally collapse faster than the cloud as a whole (Heitsch, Hartmann,
\& Burkert 2008).

We adopt an extremely simplified version of this cloud formation model;  
specifically, we use an initially circular isothermal sheet with many thermal Jeans masses
initially in hydrostatic equilibrium in the short dimension.  
We then introduce local Jeans-unstable mass concentrations in 
a spatially-random pattern within a given radius which rapidly form
sink particles (protostars).  For simplicity we do not introduce
initial velocity perturbations; instead, we allow the cloud and sinks to evolve under
their own gravity.  The random placement of the sinks (along with any density
fluctuations imposed in the gas) quickly results in complex ``turbulent''
gas velocities which are gravitationally-generated.  This setup allows us to
avoid the issue of fragmentation for the present and concentrate on the
development of CA in an initially non-clustered environment 
with a minimum of free parameters.

We use Gadget-2 \citep{Springel01, Springel05} to simulate the gas 
dynamics and the formation of ``protostellar'' sink particles. 
\citet{Jappsen05} implemented the sink particle formulation 
into the form of Gadget-2 we use. 
Collapsing structures above a density threshold ($n =10^7 \rm{cm}^{-3}$ 
in our case) are replaced 
by sink particles, which interact with gas and 
other sink particles through only gravity. 

For simplicity, we assume an isothermal equation of 
state at 10 K for the gas particles, with a molecular weight of
$\mu = 2.36$.  We use a code unit system in which the 
unit length is 1 pc, the unit time is 1 Myr and the
unit mass is 0.058 \Msun. In these units the radius of the 
sheet is then 2 pc and the total mass of the
sheet is 820 \Msun.  The surface density of the unperturbed
sheet is $1.37\times 10^{-2} {\rm g\, cm^{-2}}$ 
($A_V = 3.8$ perpendicular to the sheet). The (initial) number 
of gas particles in each simulation is $N_{tot} = 1.6\times 10^6$. 
For convenience we report results scaled to the above physical
units, but note that the simulations can be
rescaled given the assumed isothermal equation of 
state.  Specifically, if the unit length is scaled to
$d$~pc, the unit of time becomes $d$ Myr and the unit of mass 
becomes $0.058 \, d\,  \msun$.

The initial vertical structure of the sheet follows
\begin{equation}
\displaystyle{\rho(z)=\rho_0 \ \rm{sech}^2 (z/H)},
\end{equation}
with $\rho_0 = 3.7 \times 10^{-20} $ g cm$^{-3}$ and 
scale height $H = 0.06$ pc. However, the equilibrium density
distribution of an isothermal infinite sheet will follow the same form, with
a scale height of $H = c_s^2(\pi G \Sigma)^{-1} = 0.04$ pc. 

In x and y directions, the gas particles are randomly placed in a uniform 
sheet, with a radius of 2 pc (except for the non-uniform sheet case, see Section~3.2).
This leads to density fluctuations due to the random positioning 
of the particles. 
To plot the surface density and velocity fields, we interpolated the
densities and velocities of the SPH particles onto a rectangular grid.
Each cell has an area of $(0.015)^2$ in code unit or $(0.015 \rm{pc})^2$.

We start each simulation with 100 Jeans unstable clumps.
The rapid collapse of these clumps leads to 
dynamic creation of sink particles before 0.1 Myr. 
We were unable to put sinks in at the start, probably
because of problems with the boundary conditions
around the sinks;
when the sinks are dynamically created within the 
simulation, the boundary conditions are properly 
calculated to account for the discontinuities 
in density and gas pressure around the sinks \citep{Bate95,Jappsen05}. 

Because the sheet itself is also highly Jeans unstable, it 
also collapses under gravity, on a timescale
$t_c \sim R(\pi G \Sigma)^{-1/2} \approx 1.4 Myr$ (\citealt{BH04}; 
hereafter BH04).  
Due to gravitational focusing, a ring of 
material piles up quickly along the edge of the cloud. 
The edge can then become gravitationally unstable and 
fragment (BH04; \citealt{Vazquez07}; Figure~\ref{collapse}). 
With our isothermal equation of state, we find relatively
uncontrolled (numerically) fragmentation in this ring; we therefore
turn off the creation of sinks after the initial 100 clumps
collapse, allowing us to focus entirely on competitive accretion
within the main body of the cloud.  Our restriction on the initial
placement of clumps to a radius of 1 pc avoids accretion from the ring.

Gas particles that come within a certain radius of a sink (0.003 pc
in our setup) are tested for accretion individually. If a gas particle is bound
to a sink, the gas particle is accreted by the sink. 
Gas particles which come within 0.0003 pc of the sink are always accreted.
We ran each simulation for 1.2 Myr, or approximately 0.8 $t_c$,
with an output file written every 0.1 Myr.

Within this general setup we considered several cases.
In the first set of simulations, we assumed a uniform
surface density for the cloud and that each clump had
the same mass, 
0.82\Msun.  
In a second set, we assumed
the same equal initial clump masses, but a varying density
distribution in the gas.  The final sets of simulations
assumed constant surface density gas but log-normal 
initial mass distributions for the clumps, keeping the total mass of the clumps
to be 10\% of the cloud mass.
To improve statistics, we ran six realizations of each of the simulations 
described above, differing only in the random
positions of the clumps.

\section{Results}

\subsection{Equal Mass Clumps in a Uniform Sheet}

Figure~\ref{collapse} shows one of the realizations of 
the simplest case, equal mass clumps in a uniform 
sheet. The left panel shows the view from the top, 
and the right panel shows the side view. Figure~
\ref{collapse_blowup} shows a close-up view of the 
central 1.2 x 1.2 pc. The circles mark the location of the sink 
particles, and the area of the circles correspond to 
the mass of the sinks.  

Early on (before 0.2 Myr), most sinks evolve 
independently of each other, accreting mass from the 
original clump and the environment. However, as the 
entire cloud collapses, after 0.2 Myr, the sink 
particles start to affect each other, forming small groups,
in a manner reminiscent of the simulations of Bonnell,
Bate, \& Vine (2003) \nocite{BBV03}
(see also Maschberger \etal 2010).
By 0.5 Myr, the gas between the sink particles starts to 
form a filamentary structure that resembles the ``cosmic 
web" in cosmological simulations. At this stage, part of 
the gas is accreted first onto the filament, and then from 
the filament to the sinks. The regions between the web 
become depleted of gas. As time goes on, the small 
groups collapse, creating larger groups while the sink 
particles accrete gas from the environment. The more 
massive sinks in a group can accrete mass faster, thus 
broadening the mass distribution

Figure~\ref{time_evolution} shows the growth of each sink particle as a 
function of time. Initially, all the clumps have the same 
mass, but the final sink masses span over 1.5 dex in 
mass. Note that the initial clump mass is not equal 
to the sink mass when the sinks are created because it 
takes about 0.2 to 0.4 Myr for the all 
the clump gas to fall in.

Figure~\ref{m2_equal} shows the mass accretion rate of each sink at 
intervals of 0.2 Myr, including the sink particles from
all six runs. The accretion rates of the more massive
sinks exhibit a roughly
$dM/dt \propto M(\rm{sink})^2$ behavior. As the system evolves,
the accretion rates decrease due mostly to the removal 
of gas into sinks, and the lower-mass sinks lose the competition
for material to the high-mass sinks.

As shown in Figure~\ref{position_plot}, in an initially non-clustered 
environment, the accretion rate shows no clear dependence 
on the position of the sink within the sheet. 
This is unsurprising given the uniform nature of the sheet,
although the global motions of the sheet do depend upon radius.
This is in contrast to
formation in an initially clustered environment, as described 
by Bonnell et al. (2001), where the accretion rate depends on the 
position of the sink in the cluster through the tidal lobe radius. 
While the center of the cluster is the preferred location to form 
the most massive star, the most massive stars in our 
simulations do not necessarily form in the center (though eventually
everything collapses to the center).  

The combined mass distribution of the six runs is shown 
in Figure~\ref{equal_and_turbulent}. The thin black line represents the 
initial mass of the clumps. The thick lines show the mass
distribution at 0.4, 0.6, 0.8, 1.0 and 1.2 Myrs after the 
beginning of the simulation. The distribution starts with 
a delta function, evolves into a Gaussian-like distribution 
and then develops a high-mass power-law toward the end 
of the simulation. The solid black line show a fit to the 
distribution from $10^{0.4}$ to $10^{1.15}$ \Msun 
when we terminate the simulation, or at $t = 1.2$ Myr, 
with a slope of $-2.07 \pm 0.15$.  The derived slope 
does depend modestly on the range of masses which are
fitted.  The slope and the 
fitting range in mass is tabulated in Table~\ref{slope}. 

\subsection{Equal Mass Clumps in a Non-uniform Sheet}

The setup is mostly the same as the previous case, but 
with background density fluctuations. To construct a 
varying surface density, we used the linear 
superposition of sine waves in both the x and y 
directions whose magnitude is proportional to the 
wavelength: 

$$d(x,y)= \sum_{k_x,k_y} k^{-1} \sin(k_x x+\phi_x(k_x))\sin(k_y y+\phi_y(k_y)), $$
where $d(x,y)$ is the surface density at location x, y; $k_x$ and 
$k_y$ are the wavenumbers in x and y directions; $\phi_x$ and 
$\phi_y$ are the randomly chosen phases. The $k^{-1}$ factor is used
simply to ensure that the fluctuations are mostly on large scales 
while still having noticeable effects on smaller scales. 
On the smallest scales, the density fluctuations are dominated by random 
positioning of the particles. The largest wavelength allowed is the 
diameter of the sheet; the smallest wavelength allowed is $1/20$ 
of the diameter. The fluctuating part of the surface 
density is then added to a constant surface density part 
so that the minimum density is 30\% of the maximum 
density. The phases of the surface density are randomly 
chosen for each of the six simulations. 
Figure~\ref{map_fluctuation} shows a close-up view of the 
central 1.2 x 1.2 pc of one of the runs of this case. The fluctuations 
in the background density are not very prominent in the figure partly 
because the surface density is plotted on a log scale, and the 
clumps are dominating the density fluctuations. 

In this set of simulations, the accretion rate is again proportional to 
M(sink)$^2$ for the more massive sinks (Figure~\ref{fluctuation_m2}).
The sink mass distribution grows in a similar 
way as in the previous case, but the distribution spreads to higher masses
slightly faster. At $t = 1.2$ Myr, the linear fit to the 
distribution gives a slope of $-1.97 \pm 0.15$, with a 
fitting range of $10^{0.4}$ to $10^{1.15}$ \Msun.
Thus, including these density fluctuations in the simulation makes
little difference to the final result. 

\subsection{Clumps with an Initial Mass Distribution}
The previous results suggested that a wider initial distribution of 
masses should grow the power-law tail faster. We therefore 
constructed three sets
of simulations with initial mass distributions 
\begin{equation}
N(\log M) \propto \exp \left( -\frac{(\log M - \log M_c)^2}{2 \sigma^2}\right),
\end{equation}
where $\log (M_c/M_\odot) = -0.1$ and $\sigma =$ 0.05, 0.1 and 0.2 dex.
Figure~\ref{gauss} shows the sink mass distributions 
for these three cases. 
The thin black lines represent the initial clump mass 
distributions, and the thick lines are the mass distribution 
at t=0.4, 0.6, 0.8, 1.0 and 1.2 Myr. The wider 
distribution of initial clump masses yields faster growth 
of the high mass power law, as expected. Linear fits to the final mass 
distribution at t=1.2 Myr yield slope of -1.67$\pm$0.15, 
-1.42$\pm$0.14 and -1.03$\pm$0.16. Again, the parameters 
for the fitting are tabulated in Table~\ref{slope}.

The slope of the mass distribution depends on the spread of 
the initial clump masses. The final slope can be flatter than 
the Salpeter value of -1.35. In fact, if the mass accretion rate
grows strictly as $\dot{M} \propto M^2$, all the slopes would
approach -1 if the sinks have enough time and enough gas
to accrete (\eg, \citealt{Zinnecker82}).  Our numerical results are consistent with an
asymptotic slope of $\Gamma = 1.0$, although the statistical
errors are large enough to prevent an absolutely secure 
conclusion, even with simulations totalling 600 objects.
This emphasizes the long-standing problem of achieving 
sufficient numbers of objects, either theoretically or
observationally, to make firm statistical conclusions about
IMF slopes.

\section{Discussion}

\subsection{Accretion and clustered environments} 

Figures~\ref{m2_equal} and \ref{fluctuation_m2} 
show the main result of this paper: 
a strong tendency for $\dot{M} \propto M^2$ to develop 
at the high-mass end of the sink mass distribution, in initially
non-clustered, flat, collapsing cloud environments.
This results in a general tendency for the high-mass power
low to approach $\Gamma = 1$ asymptotically, depending upon
how much mass the sinks can accrete beyond their initial values,
as shown in Figures~\ref{equal_and_turbulent} and~\ref{gauss}.
To put this in context, we constructed a simple analytic
model where an initial Gaussian distribution of masses
is modified by accretion with $\mdot \propto M^2 =\alpha M^2$, 
where $\alpha$ is a constant.  
For an initial mass $M_0$, the mass grows as a function of time 
\begin{equation} 
M(t) = \frac{M_0}{1-\alpha M_0 t} 
\end{equation} 
\citep{Zinnecker82}.
The resulting mass grows as $M(t) \rightarrow \infty$ as 
$t \rightarrow t_{\infty} = (\alpha M_0)^{-1}$. 
Figure~\ref{Zinnecker} shows how the mass distribution grow
with time, plotted in increments of $0.16 t_\infty= (\alpha M_0)^{-1}$.
This does a suprisingly good job of
reproducing the numerical simulation results, if accretion
is stopped at differing times.  
Even though the simulation accretion rates do not scale
exactly as $M^2$, with the lower-mass
sinks accreting more slowly, this makes little difference on the
resulting mass distribution.   This comparison emphasizes that
the ``competitive'' effect in CA is not only starving the low-mass
systems at the expense of the high-mass objects; in terms of producing
the high-mass power-law, it is the result of differential accretion,
enhancing the rates at which the higher-mass sinks accrete.

While our starting conditions do not assume an initial clustered structure
or a deep central gravitational potential,
our assumed cloud symmetry and lack of turbulence or rotation results in
forming a cluster of sinks at the center.  However, the high-mass 
tail of the mass function is
strongly developing well before the final central cluster is formed.
Indeed, we observe $\mdot \propto M^2$ 
at the earliest stages in our simulations, where the clustering is
minimal (we also see this in a simulation with sinks
in a uniform sphere - unsurprisingly).  It does appear that
some local grouping is necessary to achieve enough differential accretion
to develop a clearly asymmetric mass function,
based on simulations (not presented here) that show when
the sinks are initially placed further apart, the groups take longer to form
and the high-mass tail of the IMF evolves more slowly.

In our simulations, the local groupings happen relatively quickly compared to 
the simulation of \citet{BCB01}. This is probably because the relaxation time
in a sheet be faster than in a sphere of the same central density and total mass
(\eg, \citealt{Rybicki72}).

\subsection{Applicability of Bondi-Hoyle accretion} 

From their simulations of formation in a cluster potential,
Bonnell \etal (2001a,b) argued that there are two regimes of accretion.
The first phase was where the gravitational potential of the cluster
gas dominated, and accretion was tidally limited, leading to a
$\Gamma \sim -0.5$.  This occurs when both the protostars and the
gas both fall in toward the cluster center (see, e.g., discussion
in Clark \etal 2009, \S 2).  During the second phase, the stars dominate
the potential, become virialized, and then Bondi-Hoyle accretion
leads to an upper mass distribution $\Gamma \rightarrow 1$. 

In contrast, we find $\Gamma \rightarrow 1$ even during global
collapse, for a situation where the infall velocities tend to be larger at large
radii and the average density is roughly constant with position
(see also Burkert \& Hartmann 2004).  This occurs as the groups
begin to dominate the local gravitational potential and generate
significant relative velocities of the sinks and the infalling gas.
This may provide local environments equivalent to the global second accretion
regime of Bonnell \etal (2001b).  The tidal limiting phase is much less
important in our simulation because of the shallower gravitational potential 
gradient of the sheet, so that the characteristic Bondi-Hoyle radius of
accretion (see below) is always smaller than the tidal radius.

In the simple, isolated version of Bondi-Hoyle accretion in three dimensions,
\begin{equation}
\mdot \propto \rho R_{acc}^2 v\,,
\end{equation}
where $\rho$ is the gas density and $v$ is the (assumed supersonic) relative
velocity of the gas and sink, both averaged at the accretion radius
\begin{equation}
R_{acc} \propto G M/v^2\,.
\end{equation}
This results in the usual scaling
\begin{equation}
\mdot \propto M^2 \rho v^{-3}\,. \label{eq:bondi}
\end{equation}

Initially, we thought that in our adopted flat geometry the
accretion rates might scale as
\begin{equation}
\mdot \propto 2 \pi  \Sigma  R_{acc} v\,,
\end{equation}
where $\Sigma$ is the gas surface density of the sheet; this
would imply
\begin{equation}
\mdot \propto M \Sigma v^{-1}\,.
\end{equation}
In fact, the accretion of the sink particles is more like a 3D than a 2D flow.
This is because the accretion radius is effectively embedded in the sheet. 
In the small groups, the velocity dispersion amongst 
the sinks is about 1- 2 $\kms$. The accretion radius is then
\begin{equation}
R_{acc} = 0.08 \left(\frac{M}{10M_{\odot}}\right) 
\left(\frac{v}{1~\rm{km~s}^{-1}}\right)^{-2} \rm{pc}.
\end{equation} 
From the above equation, we conclude that for sink masses up to $10M_{\odot}$, 
the accretion radius is in general smaller than the scale height of the sheet. 
Thus the mass flow is (non-spherical) Bondi-Hoyle accretion \citep{Bondi44}.

It is worth noting that our sheets are undoubtedly much thinner than realistic
molecular clouds.  Thus, our results suggest that formation of clouds by
large scale flows, which tend to produce flattened clouds (see \S 4.3), does not
alter the basic applicability of Bondi-Hoyle accretion for the upper mass IMF
(though conceivably the results might be different in filament geometry).

It is difficult to apply the standard formula (\ref{eq:bondi})
to our numerical results because
the background medium rapidly becomes strongly perturbed.  The gas motions
are not uncorrelated with the sink velocity, as assumed in the development
leading to equation (\ref{eq:bondi}), but instead tend to be 
{\em focused} toward mass concentrations.  The local gas density distribution
is also highly perturbed, with strong, gravitationally-accelerated flows
into and along filaments.
Bonnell \etal (2001b) attempted to deal with these difficulties through
the following argument.
Consider a point mass at radius $R$ in some environment, with infall velocities
\begin{equation}
v_{rel} \propto R^{-\eta}
\end{equation}
and gas densities
\begin{equation}
\rho \propto R^{-\xi}\,.
\end{equation}
With these assumptions Bonnell \etal found
\begin{equation}
\mdot \propto g(t) M^2 R^{3\eta - \xi}\,, \label{eq:bonnell}
\end{equation}
where $g(t)$ is a function which allows for the assumed homologous
evolution of the cluster.  This analysis results in $\Gamma \rightarrow 1$
for sinks whose masses are initially uncorrelated with position;
Bonnell \etal (2001b) suggested that the slope might be steeper
if the higher-mass objects reside preferentially in the cluster center.

To see whether the densities and velocities correlate with sink mass,
we evaluate these quantities at two radii:
first, at a radius of $2GM/c_s^2$, the maximum accretion radius in the 
Bondi accretion formulation in the case where the relative velocity
between the sink and the gas is subsonic; the other at a radius of 0.024 pc, 
which is the distance sound waves can travel in 0.1 Myr (the time between snapshots).
Figure~\ref{correlation} shows scatter plots of sink masses vs. velocities relative
to the gas, gas density, surface density and surface density 
divided by $v$, with all 
properties evaluated at R=$2GM/c_s^2$ at t=0.6Myr. 
Figure~\ref{correlation_024}  shows the same plots, with gas properties evaluated at
0.024pc away from the sink. The results show that the densities and velocities
of the gas are not strongly correlated with the individual sink masses.
Therefore, the accretion rate scales as $\mdot \propto M^2$. 
This may be a result of having a group of accreting sinks experiencing
the same environment, as in the discussion leading to equation~\ref{eq:bonnell};
whatever sets the local density and flow velocity, the capture cross-section
will still scale as $M^2$.

This suggests that the important
factor is not the form of the initial density and velocity distribution
but whether the {\em global} features are uncorrelated with the
{\em individual} sink masses, as in equation (\ref{eq:bonnell}).  In this view
as long as a group of objects of differing mass
``see'' the same conditions- gas densities and velocities - their
{\em differential} accretion rates will scale as $M^2$ (the
proportionality due to the gravitational cross-section).  This only holds
for the most massive objects in each group; the low-mass sinks are
starved of material to accrete.  More generally, the absolute value of
the mass accretion rate may vary from group to group; but as long as
each group can set up an $\mdot \propto M^2$ relative accretion rate
with differing constants of proportionality, one may argue that
the summed population will still asymptotically evolve towards $\Gamma = 1$.

\subsection{Turbulence}

Most simulations of star-forming clouds invoke an imposed
turbulent velocity field, in view of the supersonic spectral
line widths observed in molecular tracers.  Our simplified approach,
in which we do not impose initial velocity fluctuations but
initial density perturbations, is motivated by
recent simulations which form turbulent star-forming clouds from large-scale
flows \citet{Heitsch06, Heitsch08a, Heitsch08b}
and \citet{Vazquez06, Vazquez07}.  These simulations found that  
while hydrodynamically-generated turbulence in the post-shock gas
dominates the cloud structure and motions in early phases, 
gravitational acceleration dominates the motions at late stages
(e.g., \citealt{Heitsch08c}).  
Similar behavior is seen in models in which the turbulence is not
continually driven but allowed to decay (e.g., \citealt{BBB03}).
Thus, the initial turbulence provides density fluctuations or ``seeds''
which then generate supersonic motions as a result of gravitational
forces in clouds with many thermal Jeans masses.  Our models
take this view to a simple extreme, where we let gravity do all
of the (supersonic) acceleration of the gas given initial density fluctuations
(our clumps).

The assumption that the largest ``turbulent'' motions 
are mostly gravitationally-driven
is an essential part of the competitive accretion picture.
Krumholtz et al.\ (2005) argued that the supersonic velocity dispersions
of molecular clouds are too large for Bondi (and thus competitive) 
accretion to be effective; however, this assumes that the 
``turbulent'' motions persist and are spatially uncorrelated with 
the accreting masses.  In contrast, even though large 
(and roughly virial) velocities develop in our simulations,
competitive accretion still operates because the motions
are largely the result of gravitational infall to groups, plus global,
spatially-correlated collapse of both the sheet gas and the sinks.
These considerations emphasize the importance of understanding the nature
of ``turbulence'' in star-forming clouds. 

\subsection{Mass functions}

Recently, there have been suggestions that the stellar IMF is not
universal; in particular, that the most massive star in a region depends
upon its richness (\citep{KW03, KW05}; also Weidner, Kroupa, \& Bonnell 2010
and references therein).  The models presented here also result in
a non-universal upper-mass IMF, with a suggestion 
that $\Gamma \sim 1$ is an asymptotic limit which is approached most
closely when the matter accreted is much larger than the initial
``seed'' mass; and thus, to some extent, the slope may correlate with
the most massive object formed.  This is difficult to ascertain observationally,
in part because of the tradeoff between upper mass slope and truncation
mass (e.g., Maschberger \& Kroupa 2009).  Using the simulations of
\citet{BBV03} and Bonnell, Clark, \& Bate (2008),  Maschberger \etal (2010) 
found global values of $\Gamma$ slightly greater than unity, and
$\Gamma \sim 0.8$ in the richest subclusters.  This may be consistent
with our findings of a correlation between slope and upper mass.

It may be worth noting two other situations in which
$\Gamma \sim 1$ mass functions are found:
dark-matter halo simulations (below the upper-mass cutoff;
e.g., Jenkins \etal 2001); and  star cluster mass distributions
(Elmegreen \& Efremov 1997; McKee \& Williams 1997; Zhang \& Fall 1997; Chandar 2009),
although some estimates yield flatter power-law slopes (e.g., 
Maschberger \& Kroupa 2009).
Gravitational accretion thus could potentially provide a unified explanation
of the similarities in these mass functions.

\section{Conclusions}

This paper presents numerical experiments using SPH simulations to address 
the general applicability of competitive accretion in initially non-clustered 
environments. A flat geometry is used to construct a shallow gravitational potential
as opposed to the spherical clustered potential used in previous simulations by 
Bonnell \etal (2001a,b). The simplified setup consists of only the most important 
elements in forming the high-mass IMF: differential gas accretion onto 
protostars under gravity.
With this setup, we were able to produce the high-mass end of 
the IMF with slopes comparable to the Salpeter slope $\Gamma = 1.35$. 
The simple setup also allows us to understand the mass growth of sinks
in details without worrying about fragmentation and thermal physics, and
also permits us to generate reasonably statistically-significant results for
upper mass function slopes.

The mass growth rate of the sinks follows $\dot{M} \propto M^2$ for all high 
mass sinks, while low mass sinks sometimes accrete at lower rates. 
The high-mass end of the IMF develops a power-law tail and flattens, with 
an asymptotic slope of $\Gamma = 1$. Variations in initial clumps masses 
and surface density help the power-law tail to flatten faster. In our simulations,
most systems do not reach the asymptotic slope due to gas depletion. In real 
molecular clouds, stellar feedback as well as gas depletion can terminate the 
accretion and determine the final high-mass IMF slope.

The present set of simulations are obviously quite idealized.  Our purpose
was to elucidate the basic physics of CA in as easily-visualized and
interpretable a situation as possible.  The next steps,
which are currently under way, are to start with 
more complex density distributions
and allow sink formation and consequent evolution in more complex geometries,
and include velocity fields as necessary.
While we suspect that the physics of competitive 
accretion will remain the most important factor in creating the 
high-mass region of the IMF, as previously
argued by Bonnell \etal (2001a,b, 2003), and Clark \etal (2009),
\nocite{BBB03,BBV03}
further study is needed.

\acknowledgments
We wish to thank the anonymous referee for very helpful comments which improved the 
paper substantially. We thank Jeremy Hallum for his efforts in maintaining the cluster on which
these simulations were computed.  This work was supported in part by
NSF grant AST-0807305 and by the University of Michigan.

\bibliography{./Bib} 
\input{slope.tex}
\input{figure.tex}

\end{document}

%% file: slope.tex
\begin{deluxetable}{cccccc}
  \tabletypesize{\small}
  \tablewidth{0pc}
  \tablecaption{Fitting range and slope of the mass function}
  \tablehead{\colhead{Case}& \colhead{Background}  & \colhead{Clump Mass}  & \colhead{Fitting Range} & \colhead{Slope} &\colhead{\% of Final Mass in Sinks} \\
   		 \colhead{}&\colhead{}&\colhead{}&\colhead{$\log$(\Msun)}& \colhead{}&\colhead{} }
  \startdata
  
	1	& uniform	&  same   &  0.4 - 1.15 & -2.07$\pm$ 0.15 &  39\% \\
    	2	& varying	&  same	&  0.4 - 1.15 & -1.97 $\pm$ 0.15&  40\% \\
	3	& uniform	& gaussian ($\sigma = 0.05$) & 0.4 - 1.15  & -1.67$\pm$ 0.14& 40\% \\
	4	& uniform	&  gaussian($\sigma = 0.1$) & 0.45 - 1.2  & -1.42 $\pm$ 0.14& 40\% \\
	5	& uniform	& gaussian($\sigma=0.2$) & 0.55 - 1.3 & -1.03 $\pm$ 0.16 & 40\% \\
    \enddata
    \label{slope}
\end{deluxetable}

%% file: figure.tex
\clearpage

\begin{figure}
\begin{center}
	\includegraphics[scale=0.38]{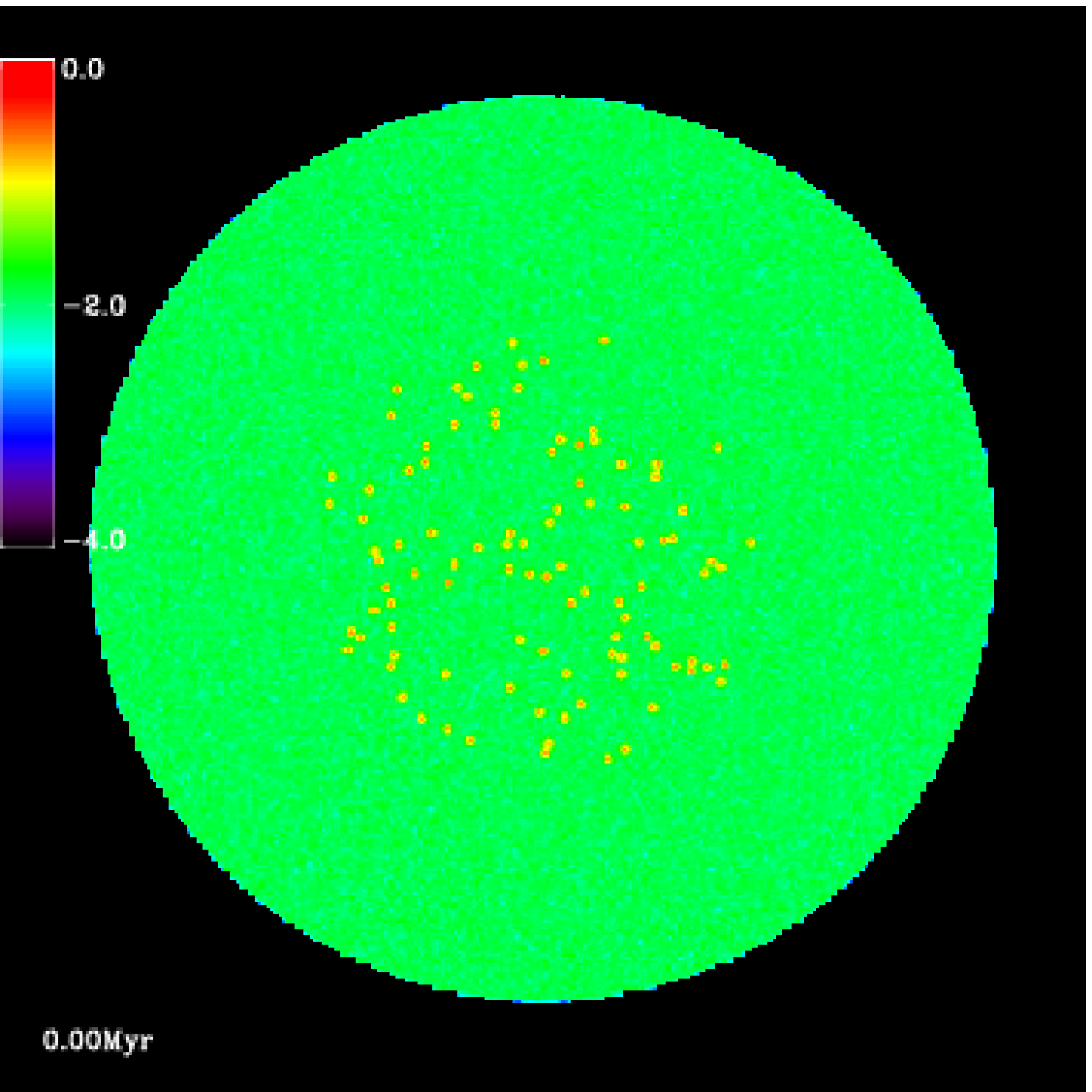}
	\includegraphics[scale=0.38]{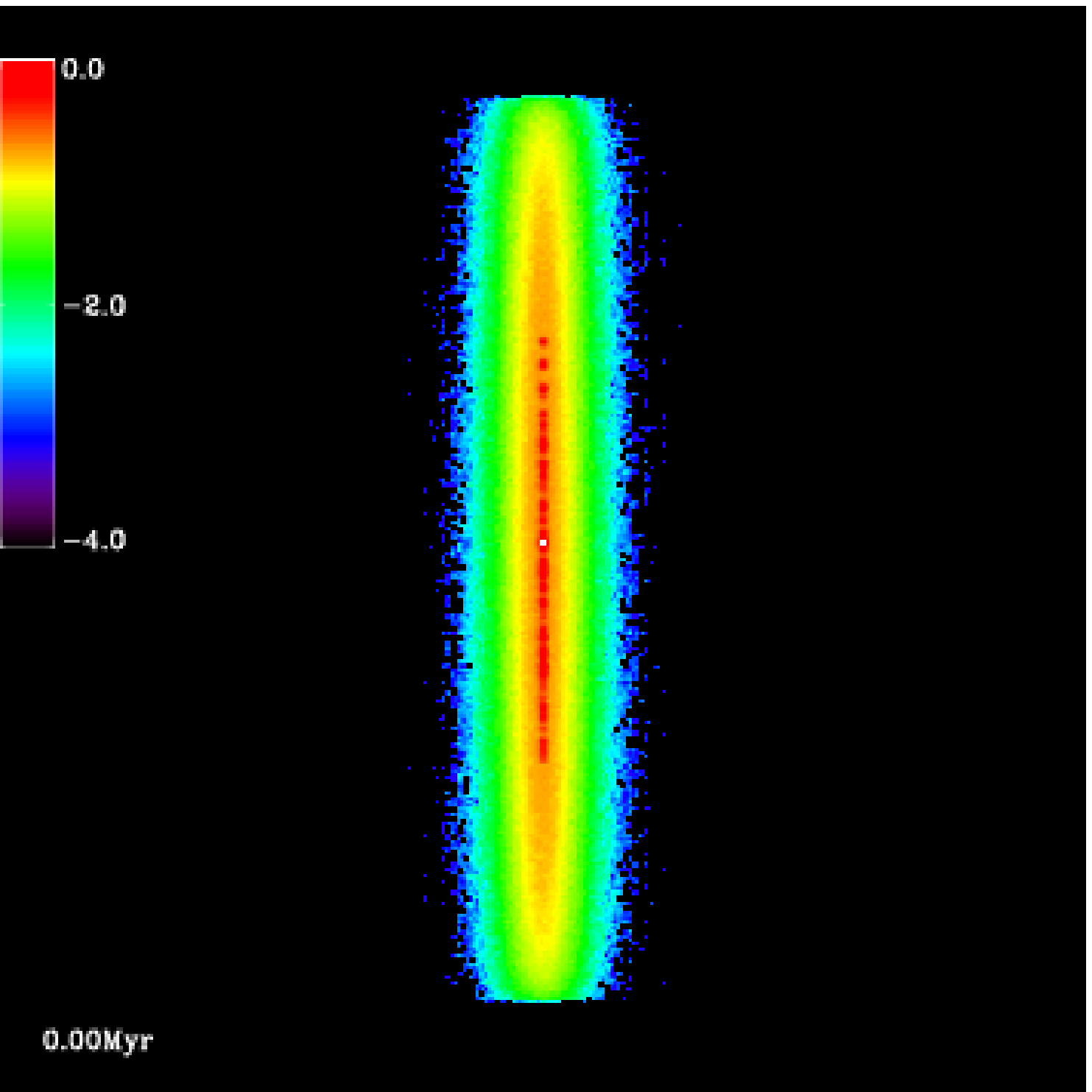}\\
	\includegraphics[scale=0.38]{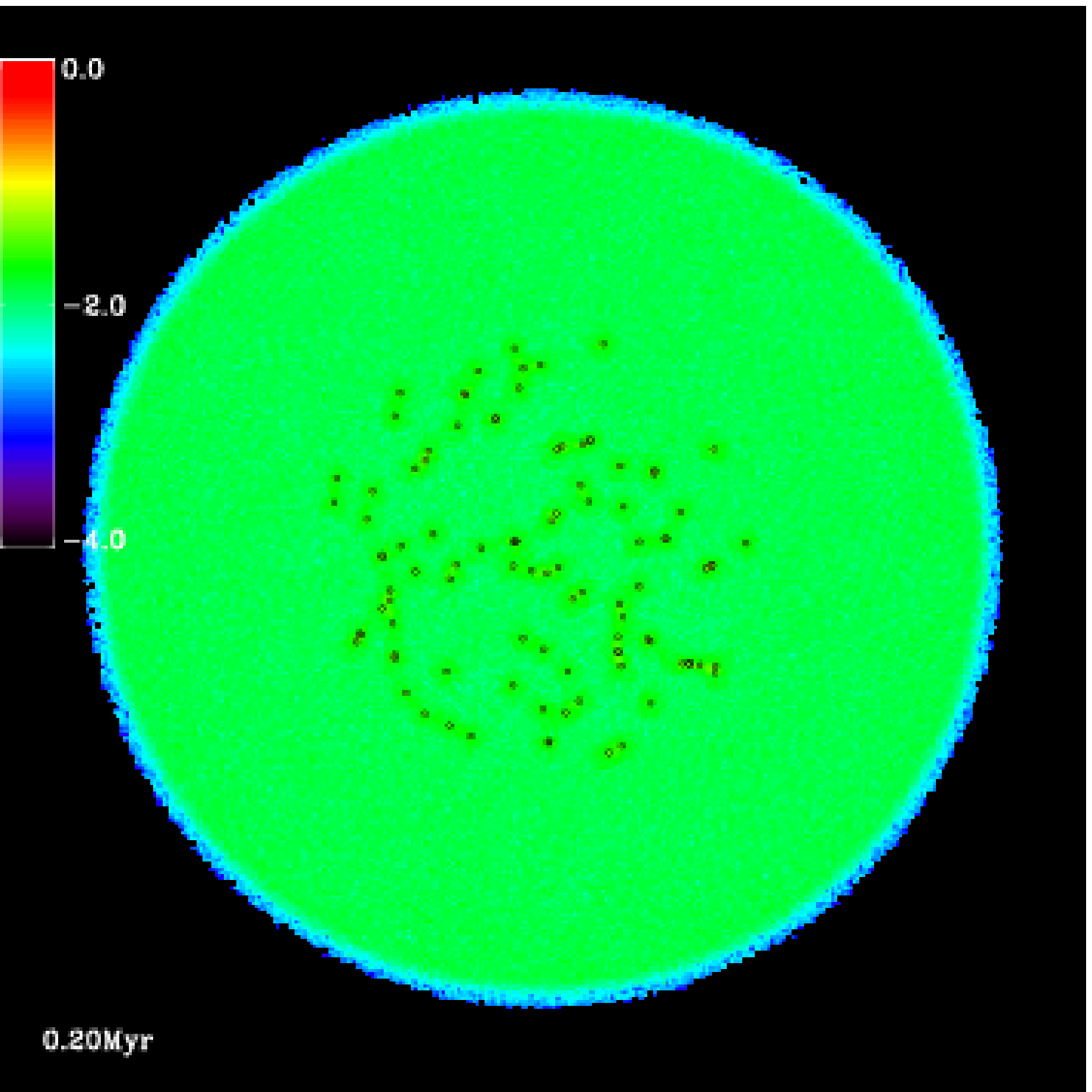}
	\includegraphics[scale=0.38]{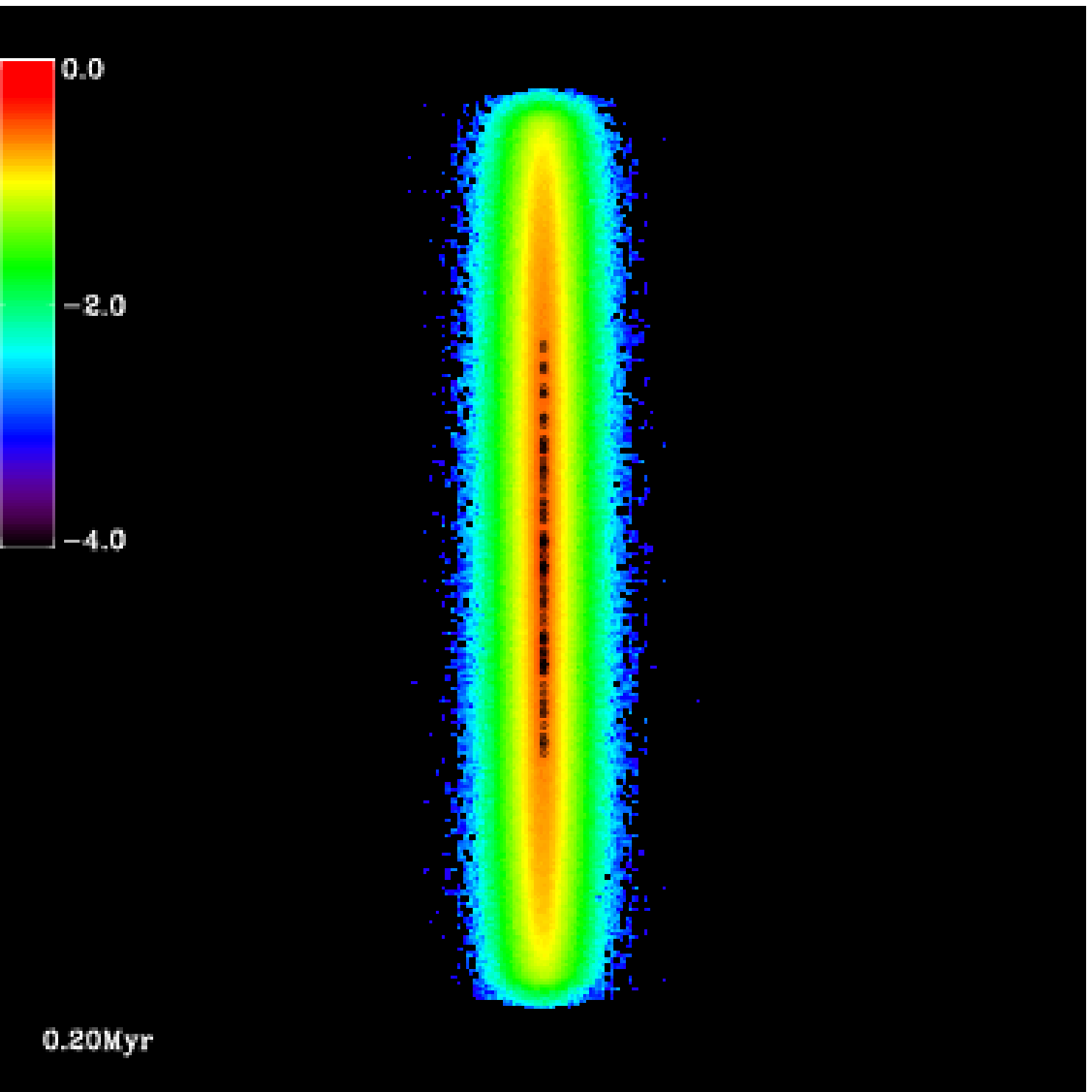}\\
	\includegraphics[scale=0.38]{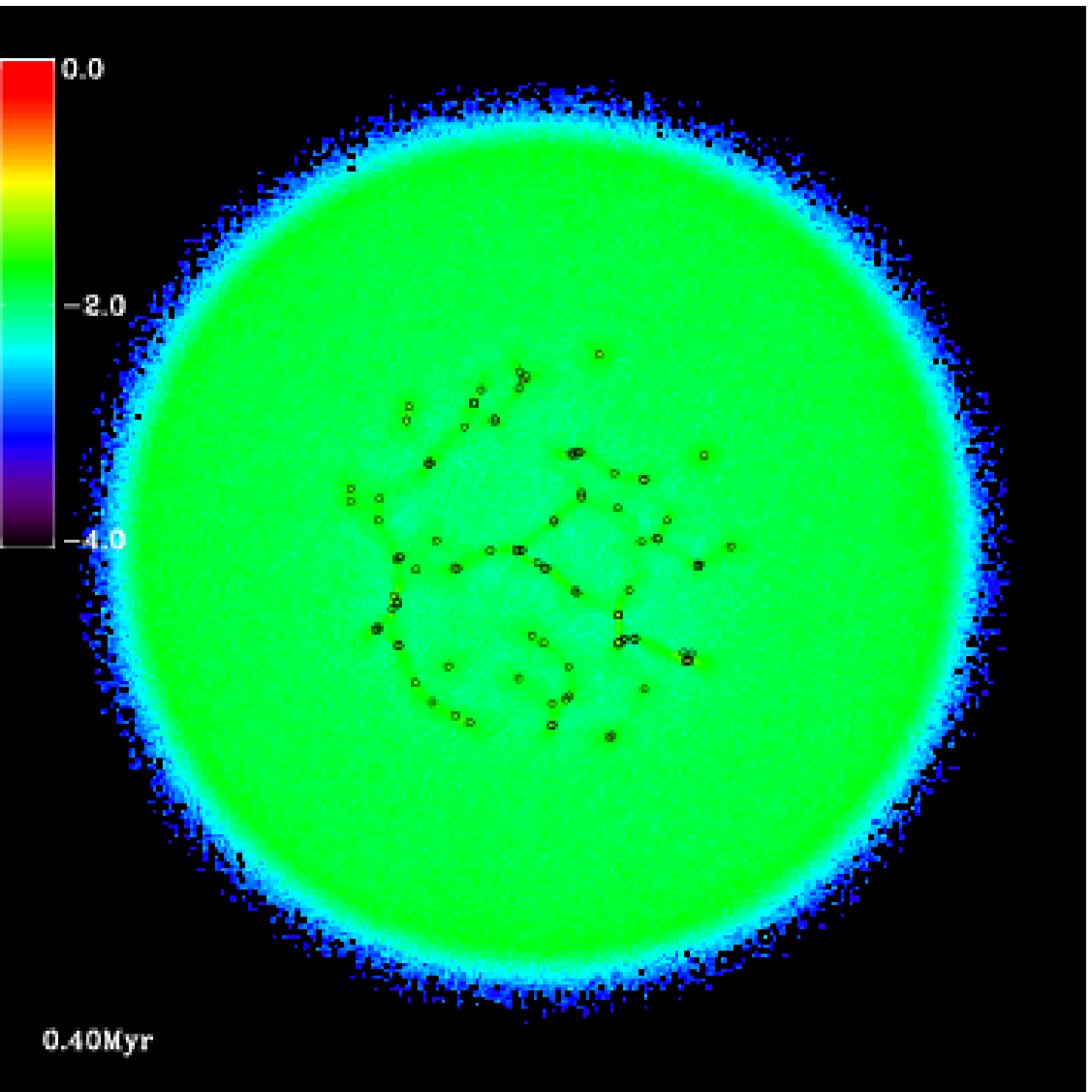}
	\includegraphics[scale=0.38]{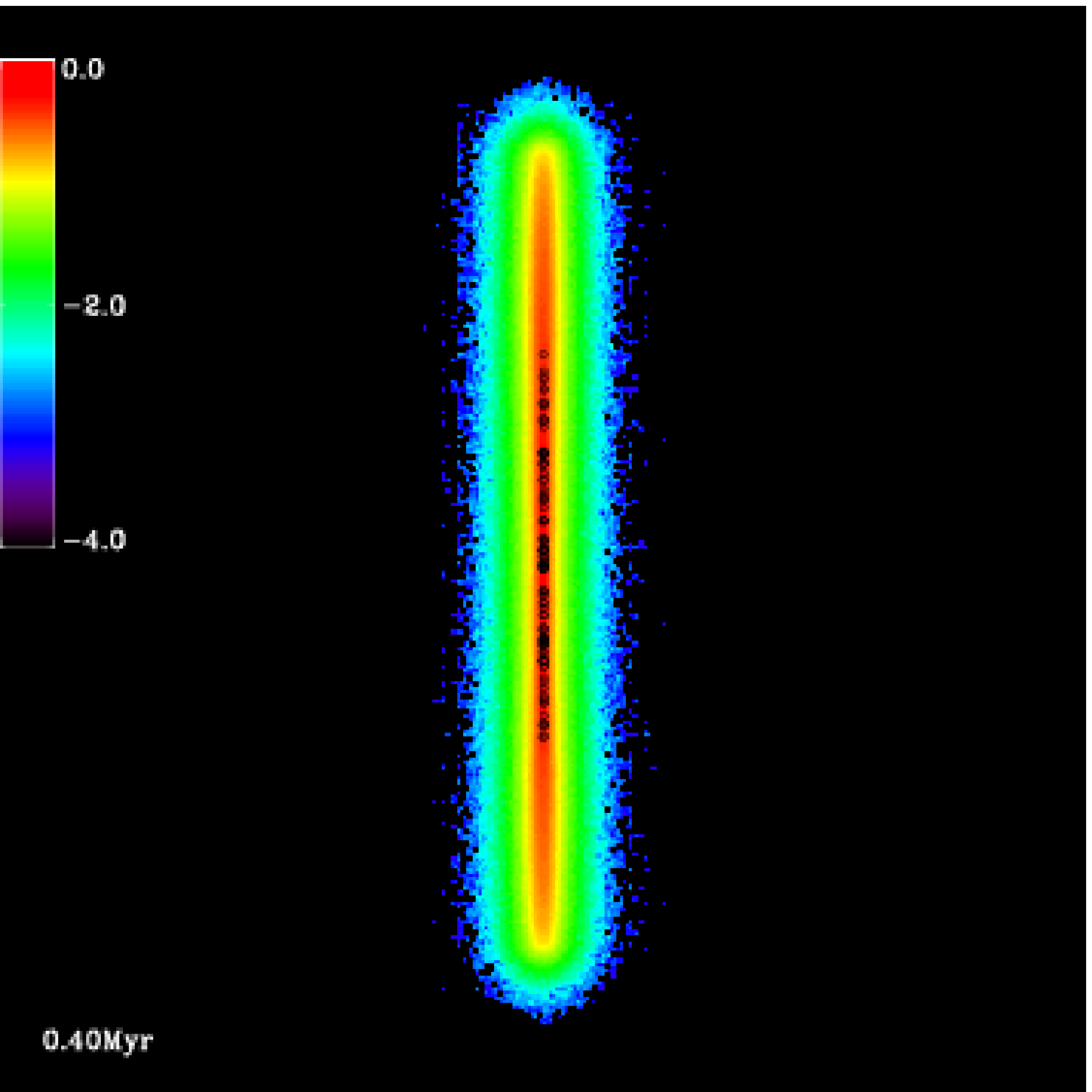}\\
	\includegraphics[scale=0.38]{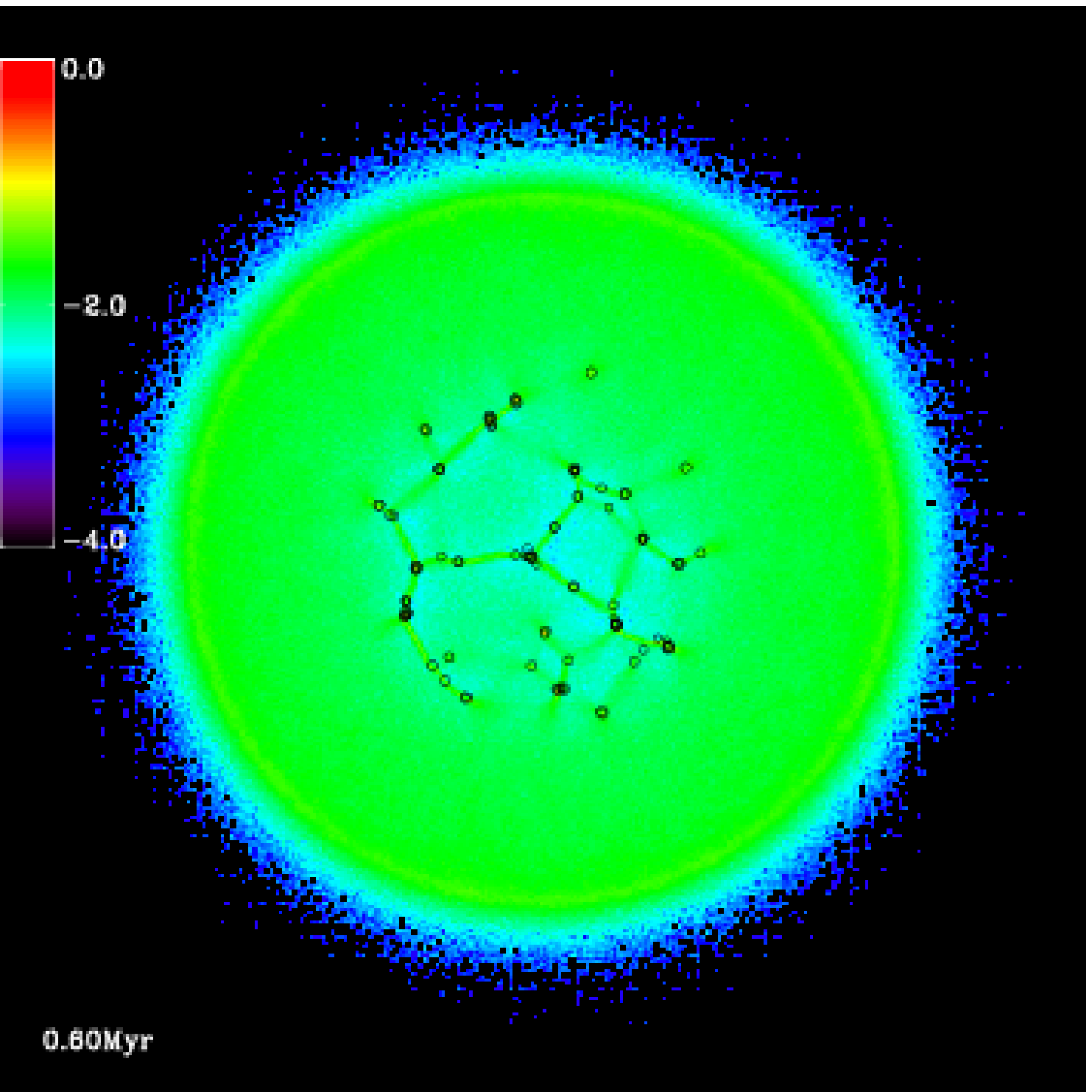}
	\includegraphics[scale=0.38]{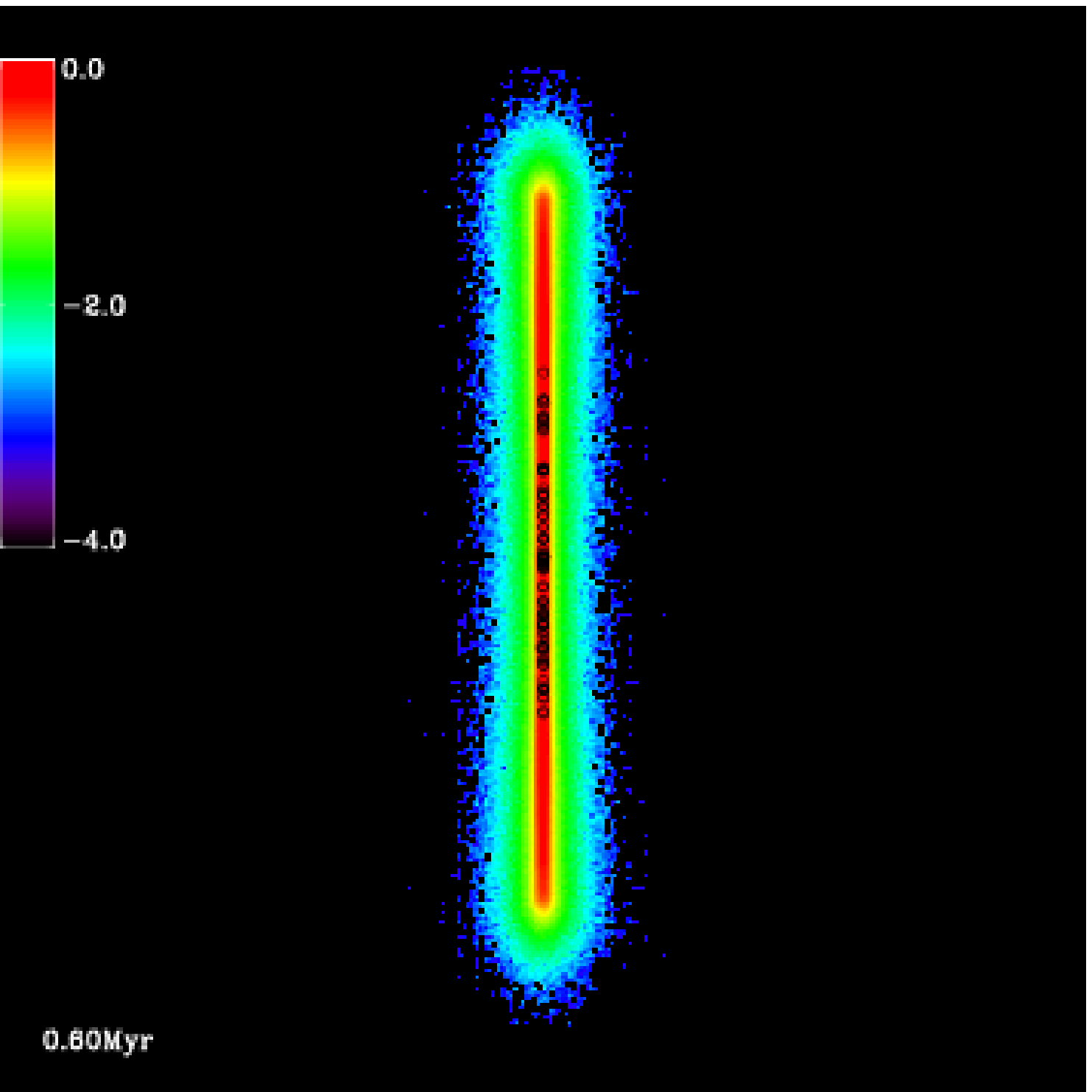}
	\end{center}
\end{figure}

\begin{figure}
	\begin{center}
	\includegraphics[scale=0.38]{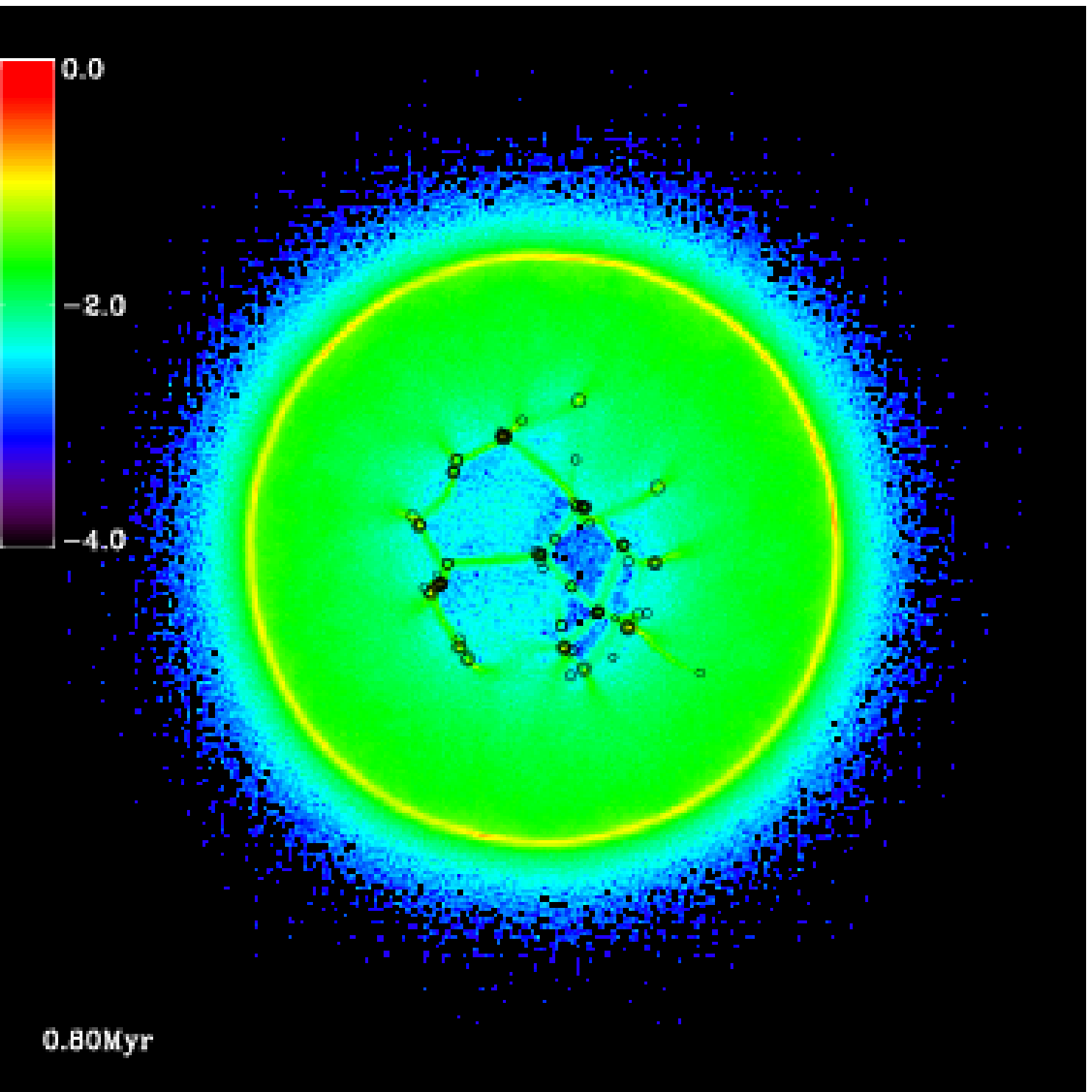}
	\includegraphics[scale=0.38]{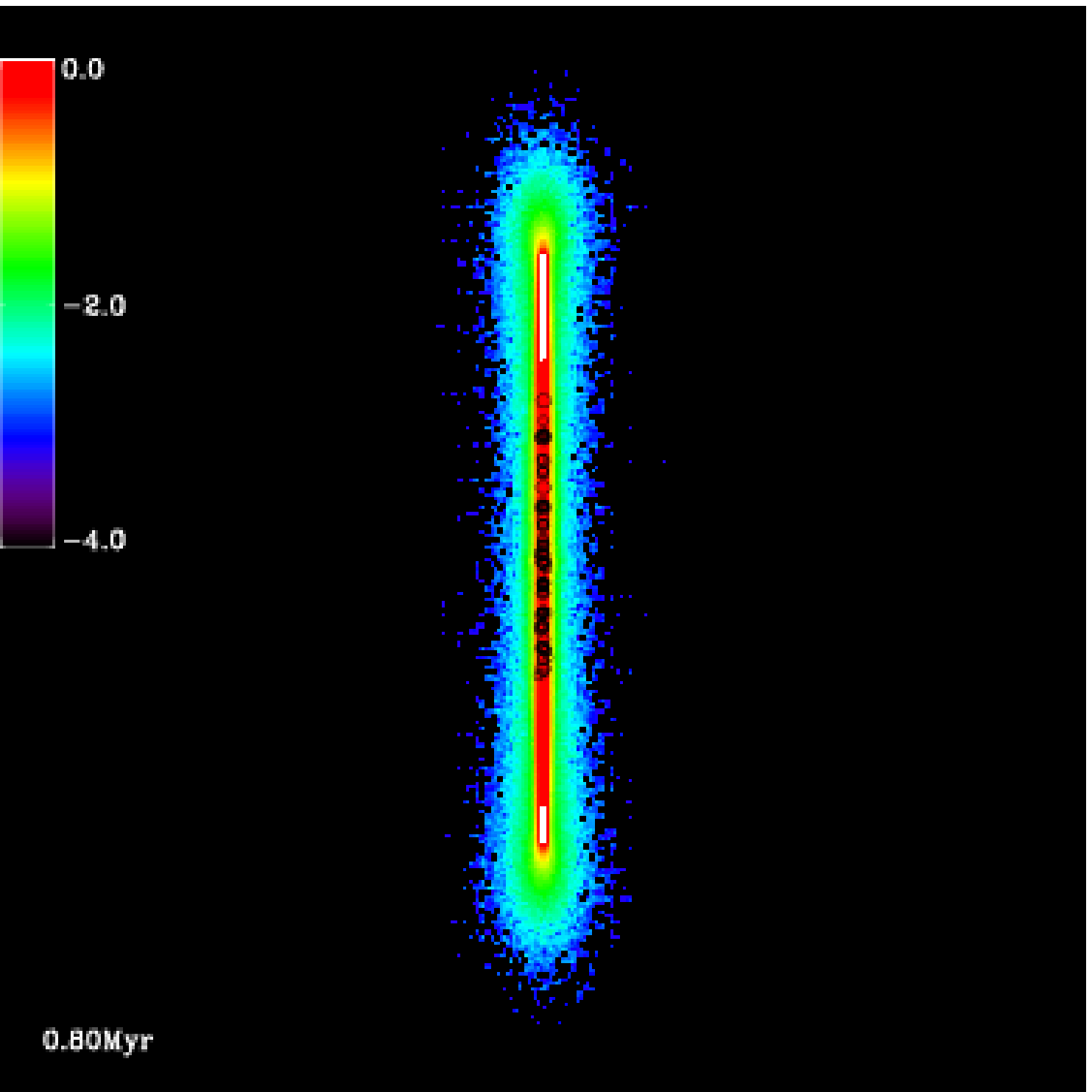}\\
	\includegraphics[scale=0.38]{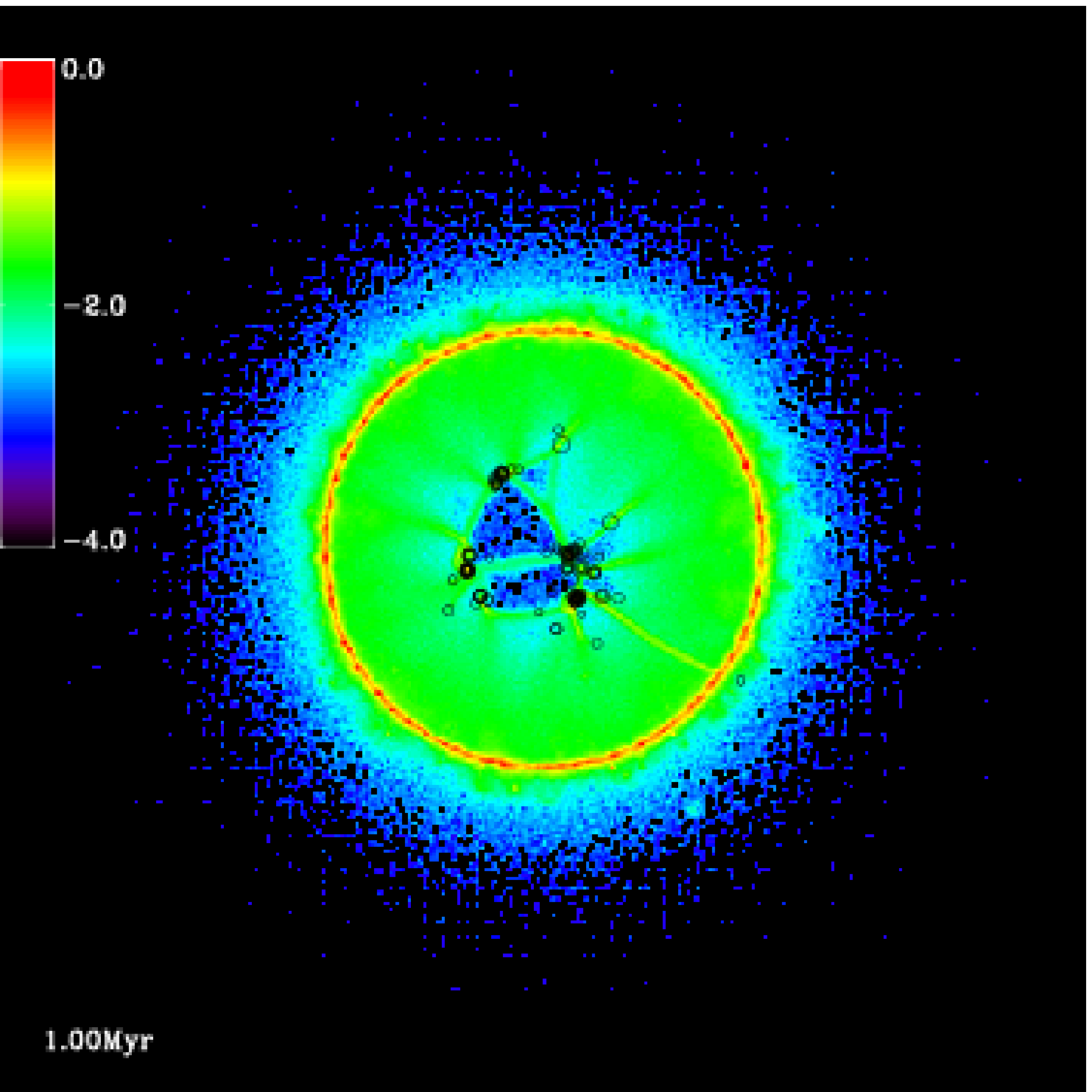}
	\includegraphics[scale=0.38]{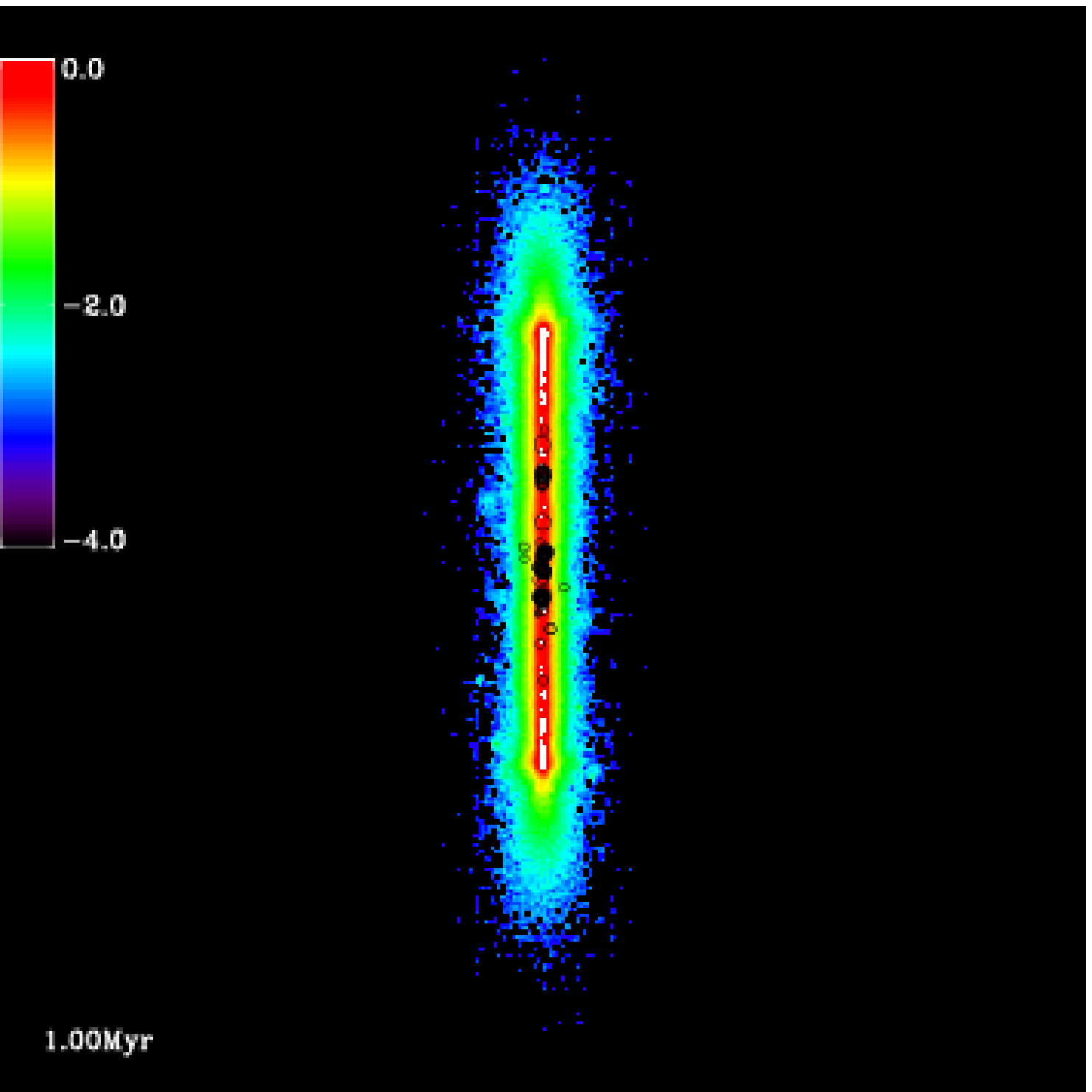}\\
	\includegraphics[scale=0.38]{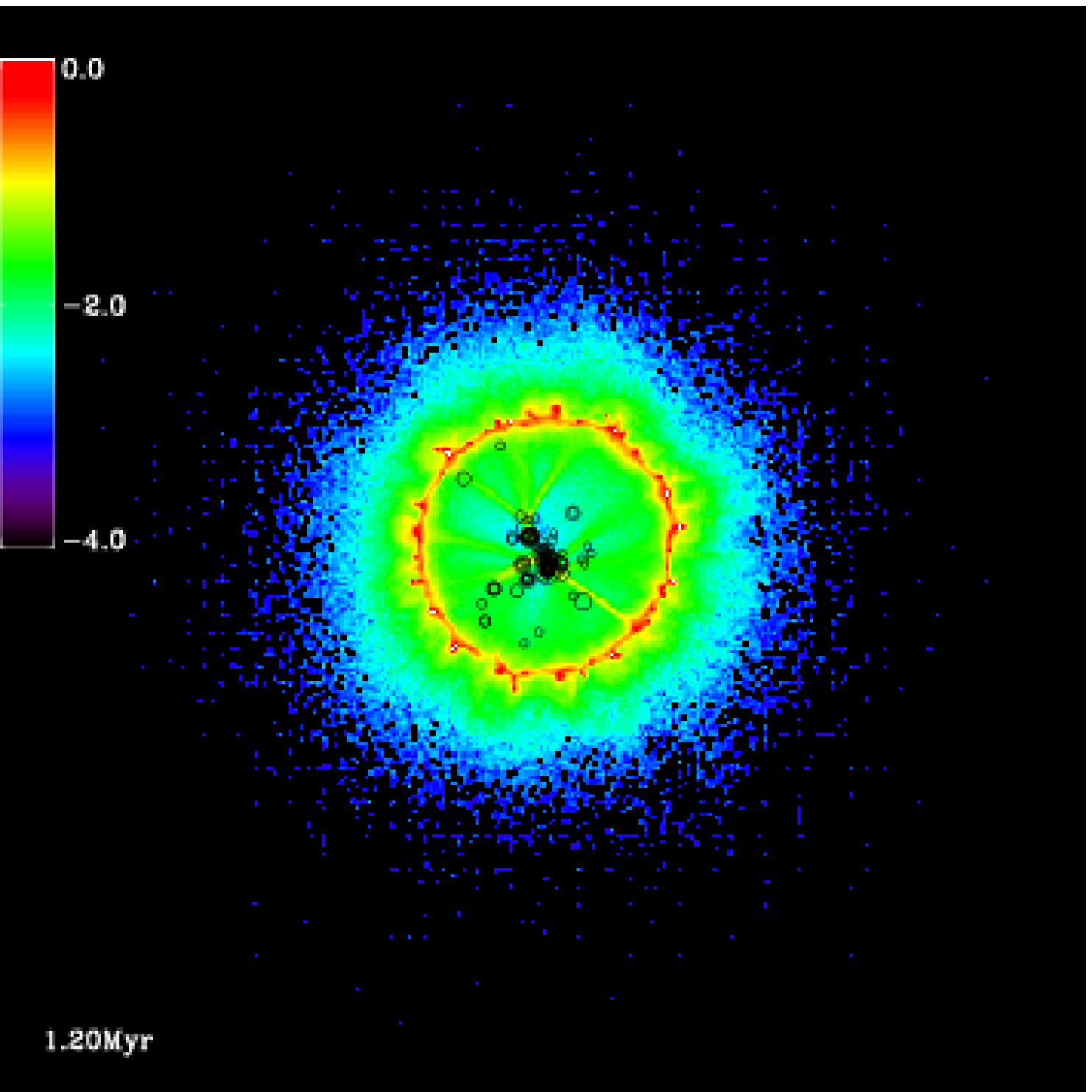}
	\includegraphics[scale=0.38]{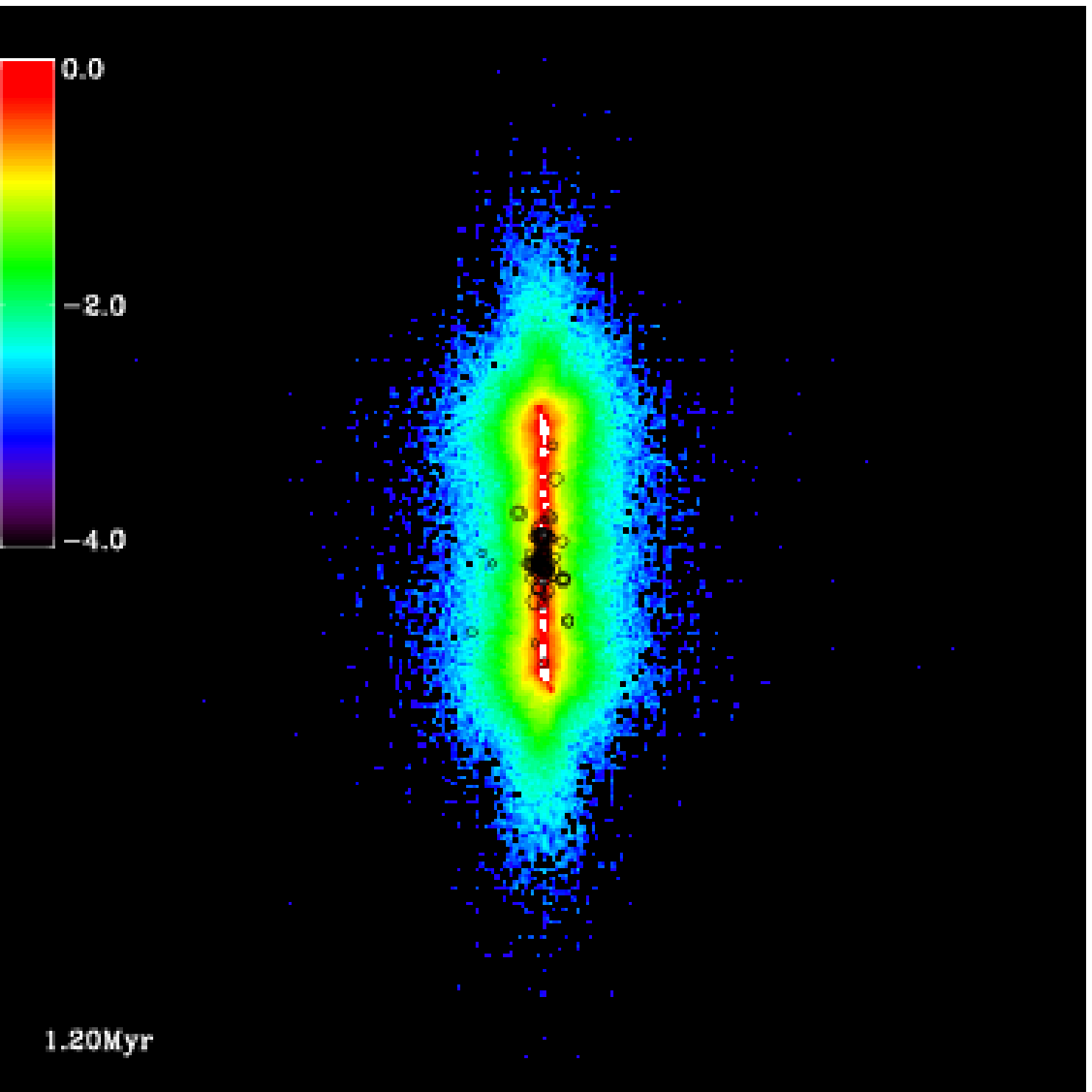}
	\caption{The collapse of a sheet-like molecular cloud and the growth of the clumps in a simulation at 0, 0.2, 0.4, 0.6, 0.8, 1.2 Mpc. The left panel is the molecular cloud as viewed from the top, and the left panel from the side. Each box is 4.8 by 4.8 pc. The colors correspond to the logarithm of column density in $\rm{g~cm}^{-2}$.}
	\label{collapse}
		\end{center}
\end{figure}

\begin{figure}
	\begin{center}
	\includegraphics[scale=0.38]{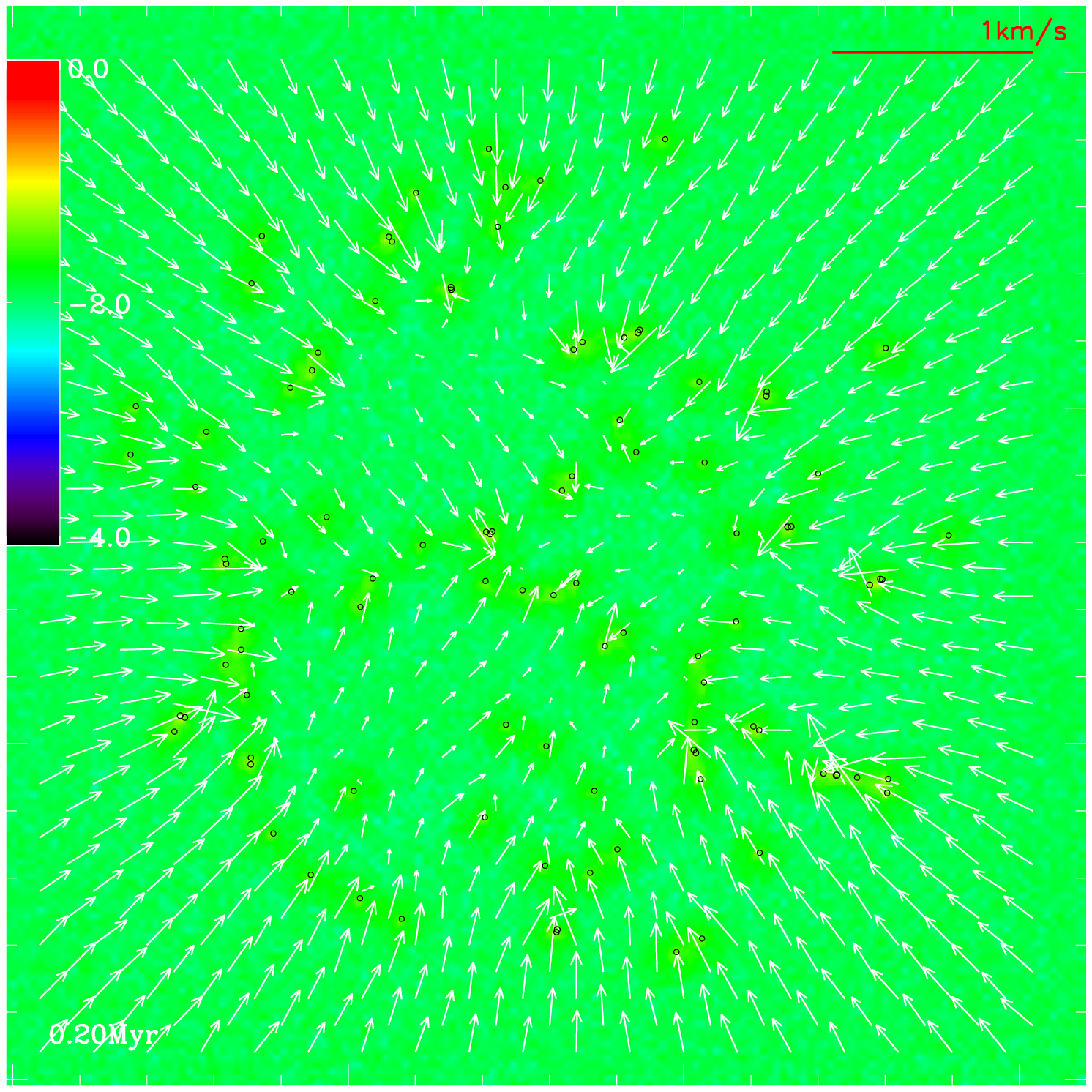}
	\includegraphics[scale=0.38]{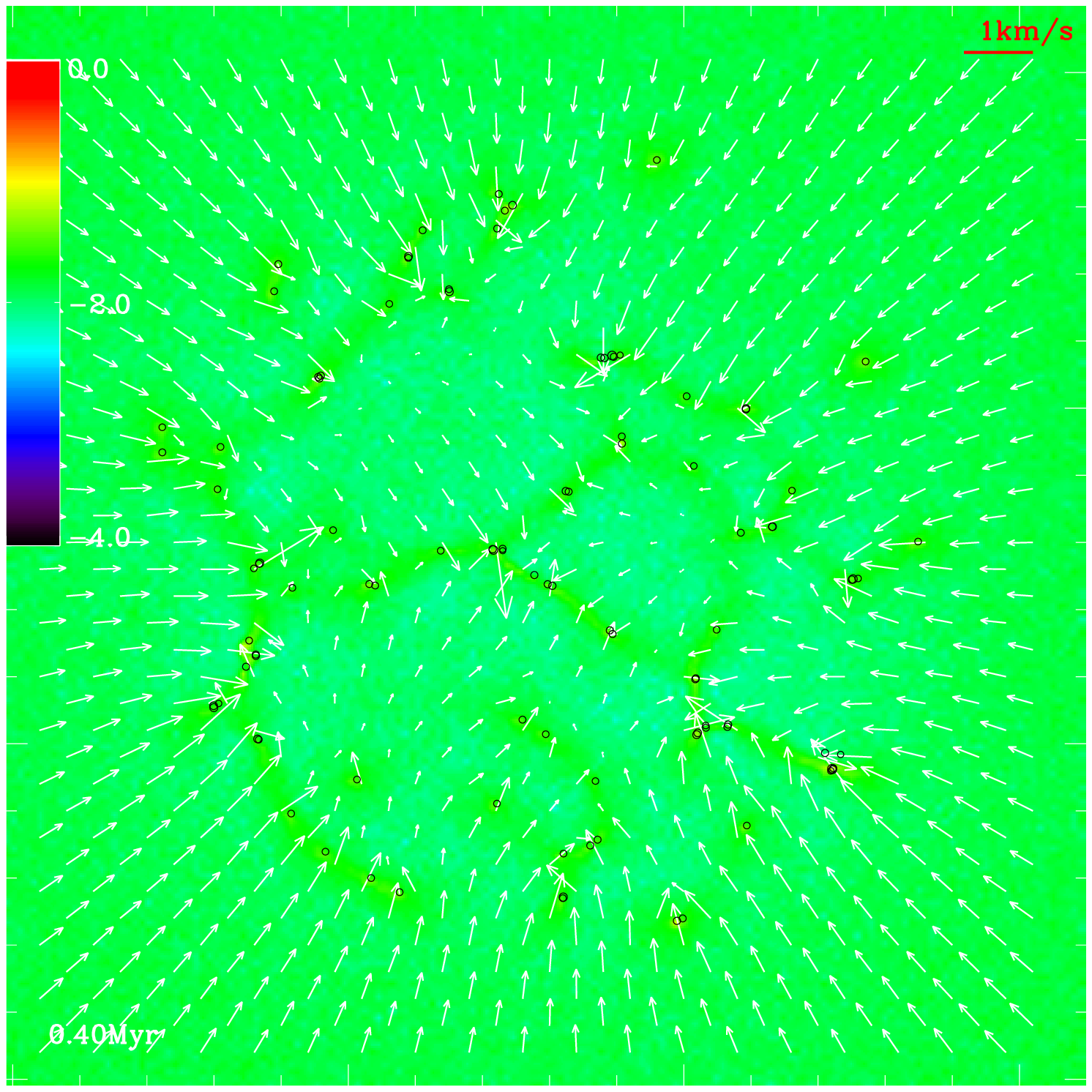}\\
	\includegraphics[scale=0.38]{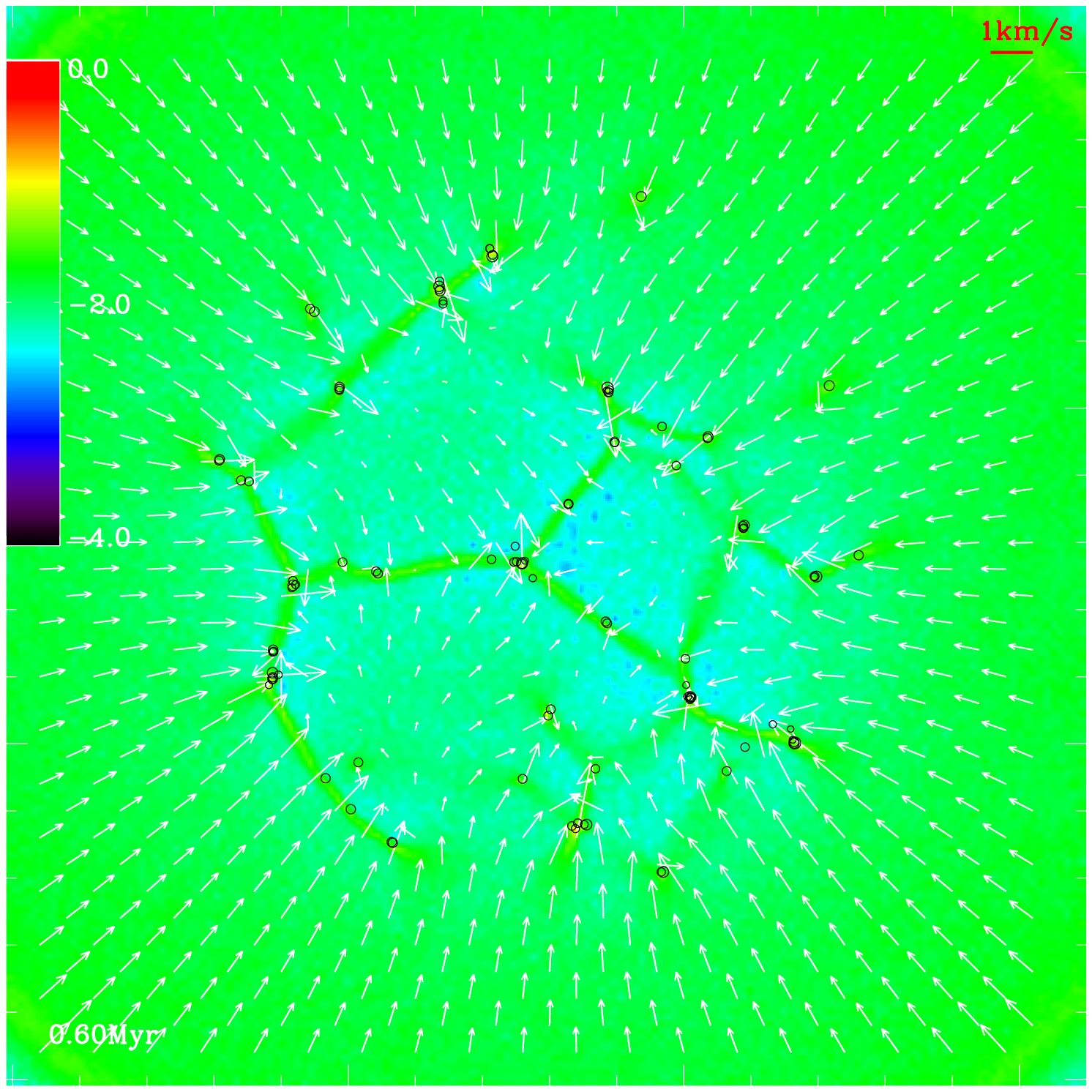}
	\includegraphics[scale=0.38]{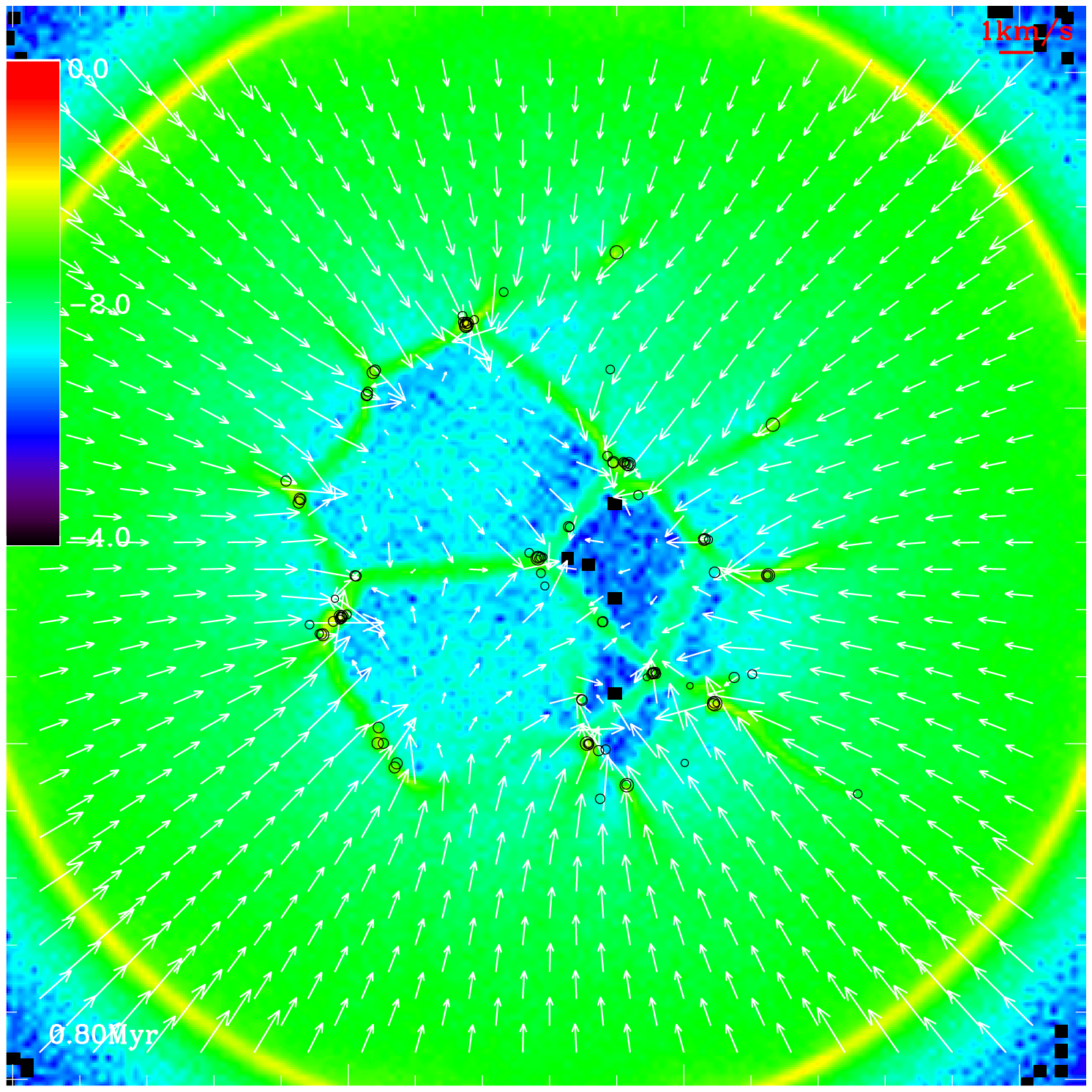}\\
	\includegraphics[scale=0.38]{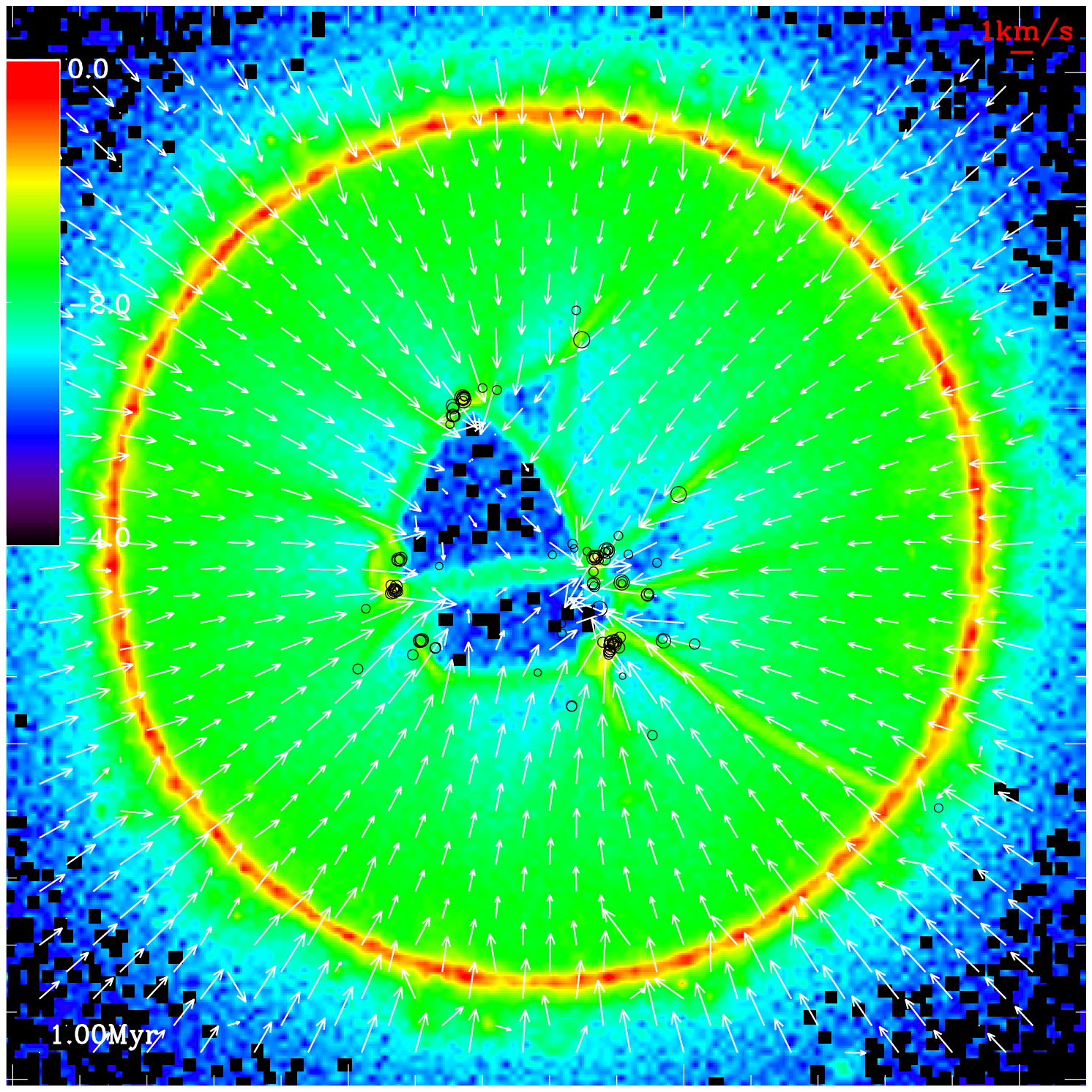}	
	\includegraphics[scale=0.38]{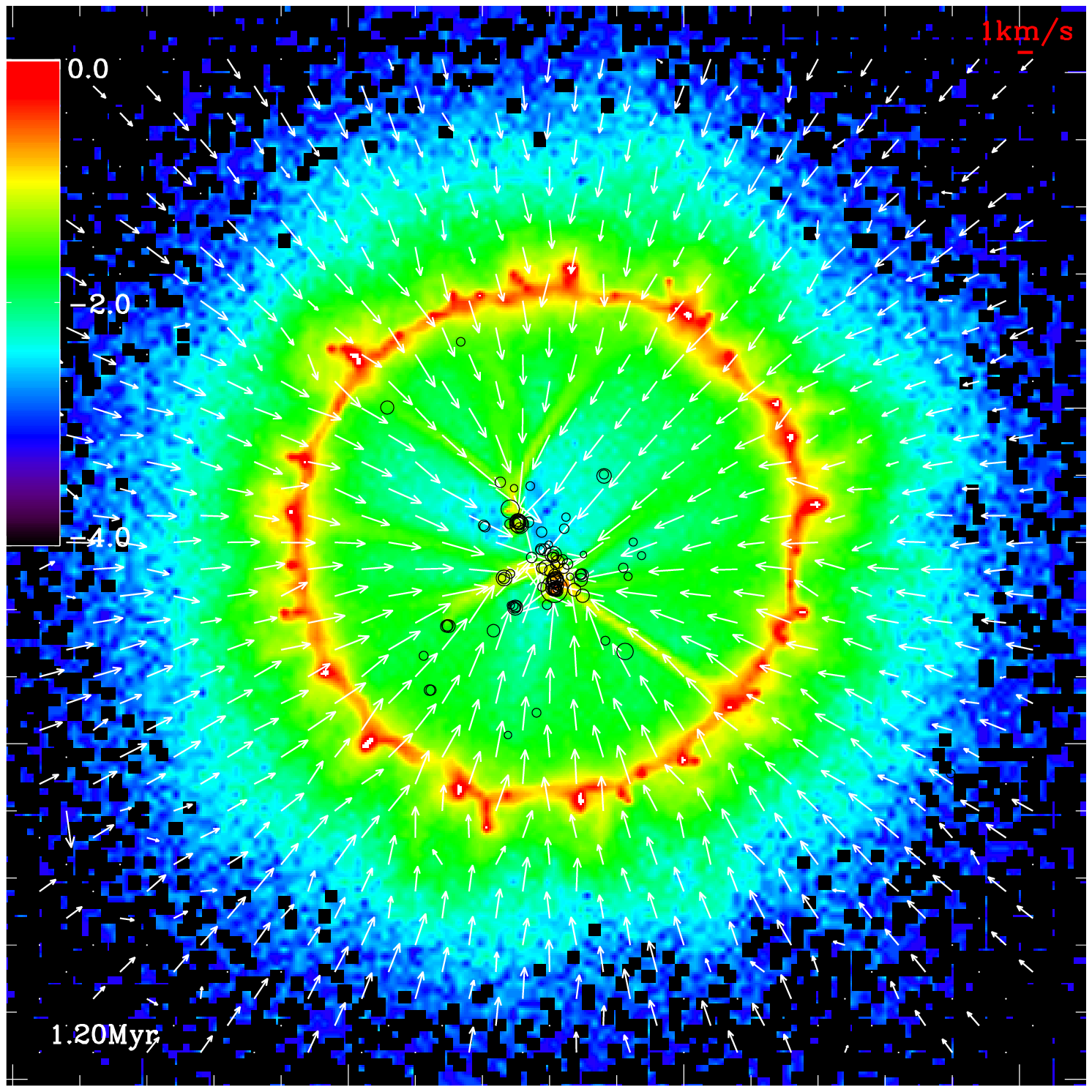}
	\caption{The same simulation as in Figure~\ref{collapse}, but only the inner part is shown. Each box is 2.4 pc by 2.4 pc. The arrows indicate the velocity vectors of the gas, with 1$\kms$ marked on the upper right corner of each panel.  The colors correspond to the log of column density in $\rm{g~cm}^{-2}$.}
	\label{collapse_blowup}
	\end{center}
\end{figure} 

\begin{figure}
	\begin{center}
	\includegraphics[scale=0.8]{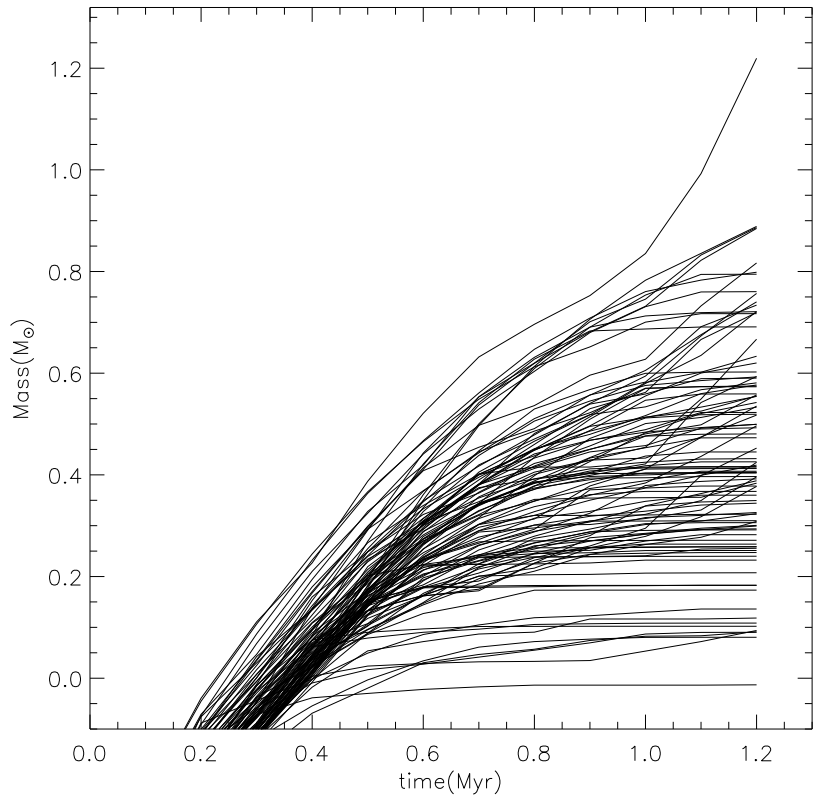}
	\caption{The evolution of sink mass as a function of time in one of the runs of the equal mass case.}
	\label{time_evolution}
	\end{center}
\end{figure}
		
\begin{figure}
	\begin{center}
	\includegraphics{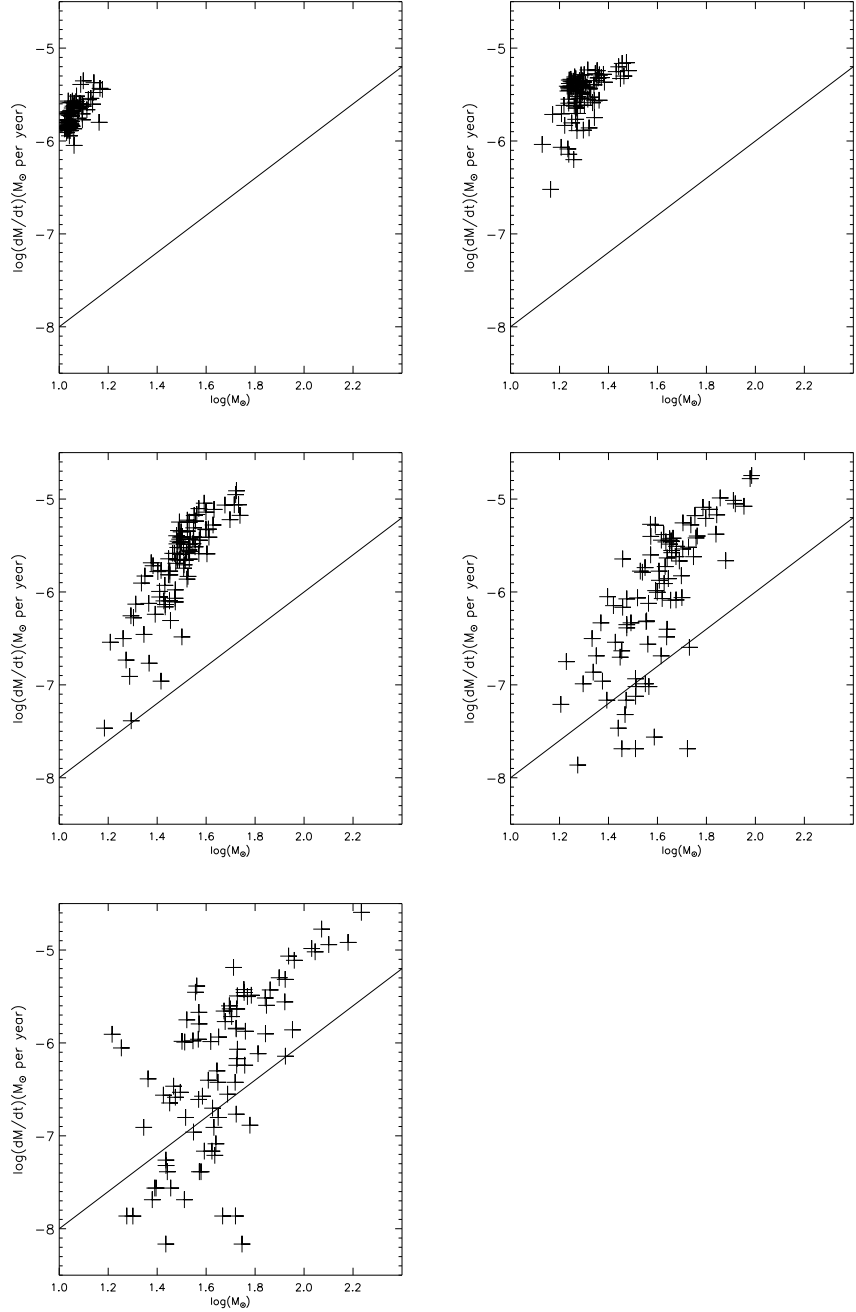}
	\caption{The accretion rate vs. mass of sinks at 0.2, 0.4, 0.6, 0.8 and 1.0 Myr in one of the runs of equal mass sinks, with the uniform surface density cloud. The accretion rates for the higher-mass sinks follow $\dot{M} \propto M^2$.}
		\label{m2_equal}
	\end{center}
\end{figure}

\begin{figure}
	\includegraphics{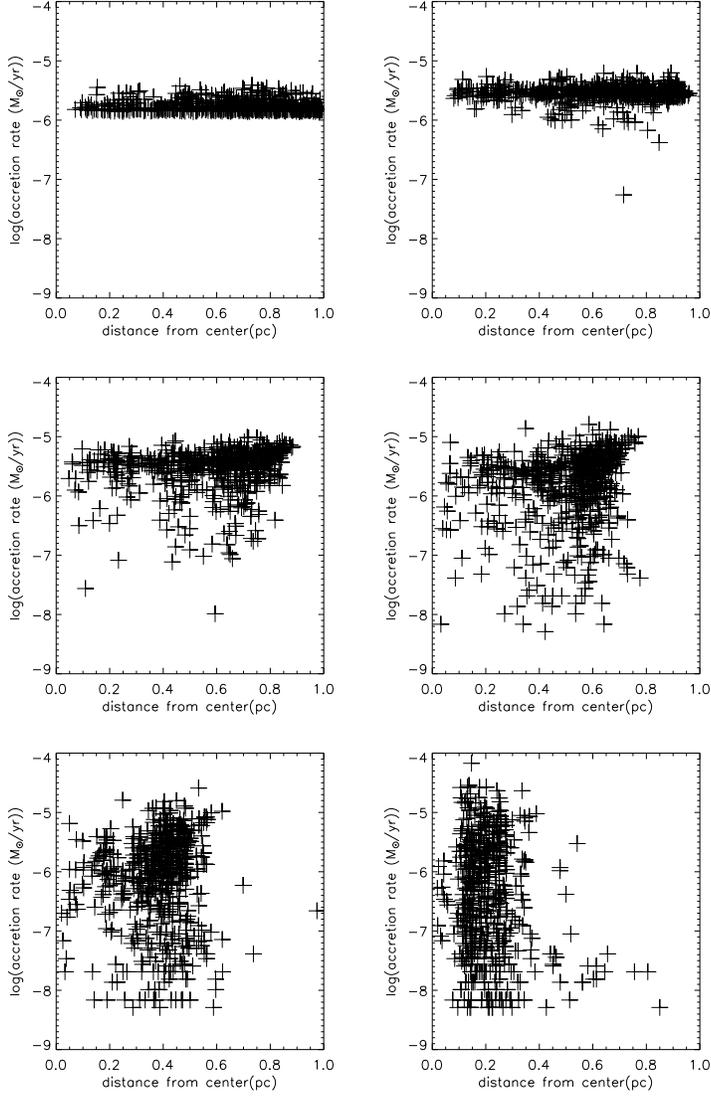}
	\caption{The accretion rate vs. distance of sink from the center of the sheet at t=0.2, 0.4, 0.6, 0.8, 1.0 and 1.2 My in the runs with equal mass sinks and the uniform surface density cloud. All 600 sinks from the six runs are included.}
		\label{position_plot}
\end{figure}

\begin{figure}
	\includegraphics[scale=0.7]{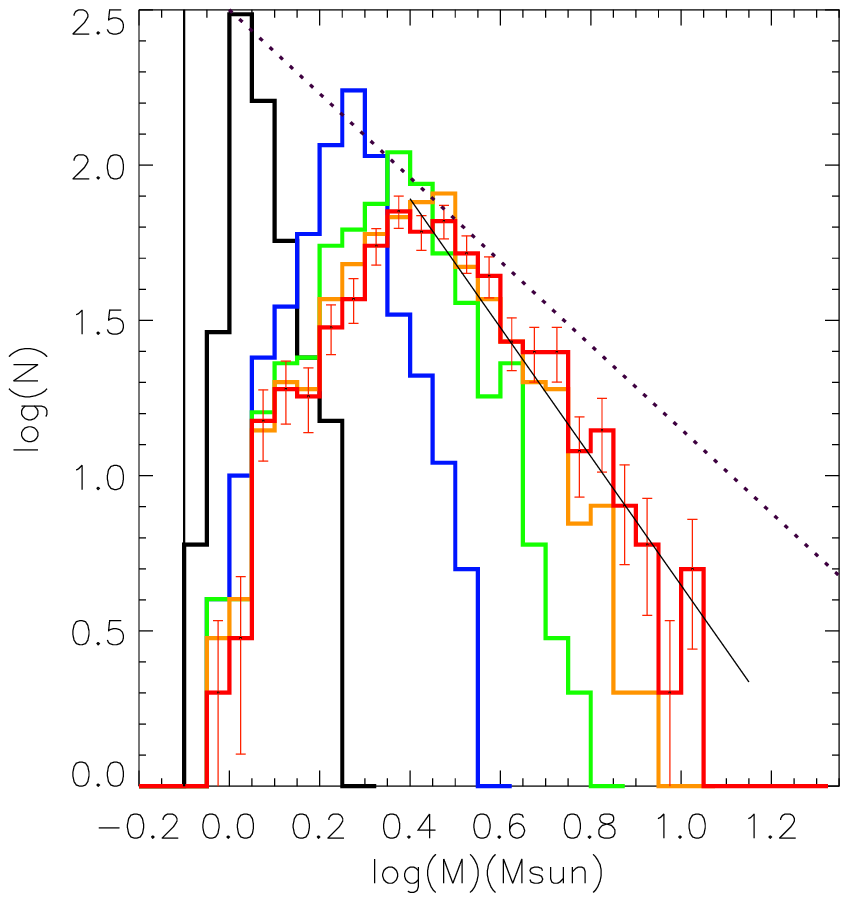}
	\includegraphics[scale=0.7]{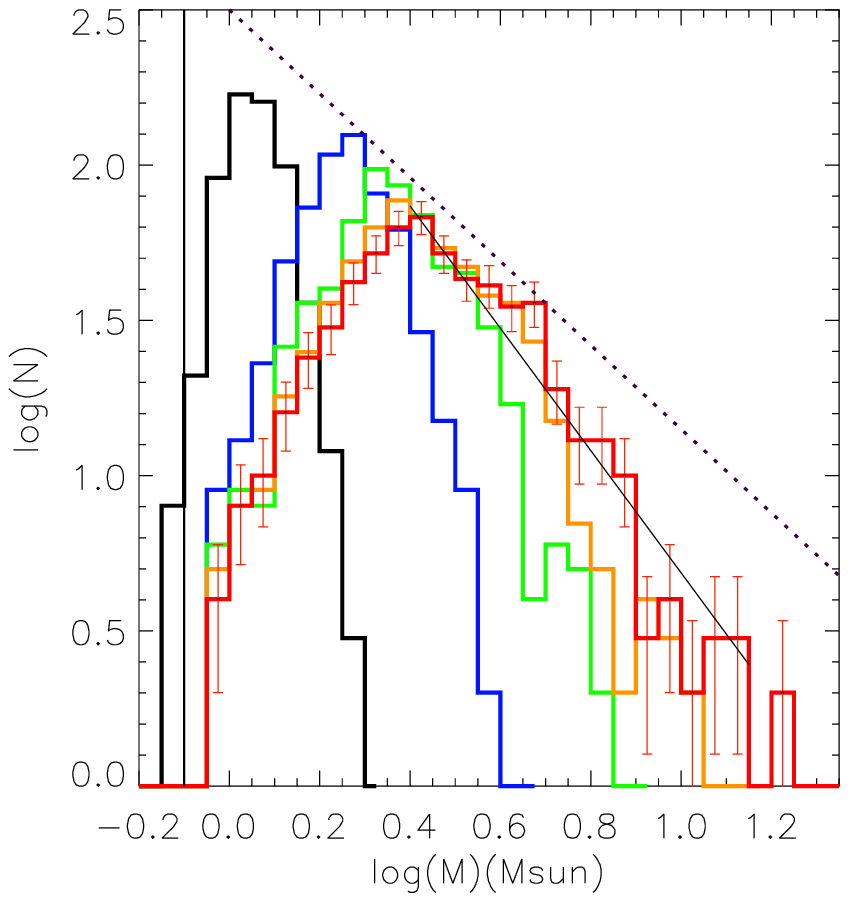}
	\caption{Left: The distribution of mass at t=0.4, 0.6, 0.8, 1.0 and 1.2 Myr for the constant background density, equal clump mass case. The weighted linear fit has a slope of -2.08. Right: The distribution of mass at t = 0.4, 0.6, 0.8, 1.0 and 1.2 Myr for the varying background density, equal clump mass case. The weighted linear fit has a slope of $-\Gamma$ = -1.95. The dotted lines represent the Salpeter slope $-\Gamma$ = -1.35 (see Table 1).}
	\label{equal_and_turbulent}
\end{figure}

\begin{figure}
	\begin{center}
	\includegraphics[scale=0.38]{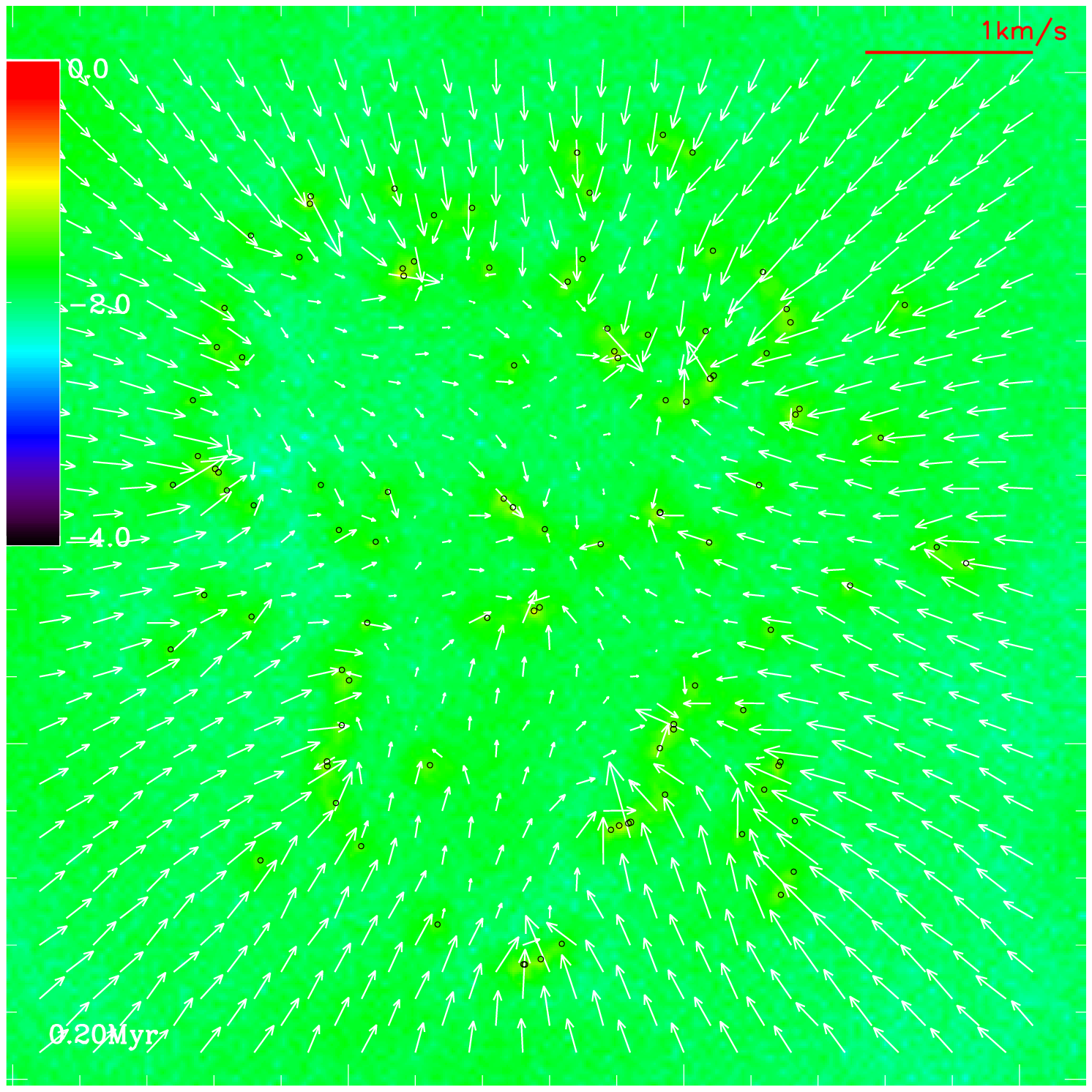}
	\includegraphics[scale=0.38]{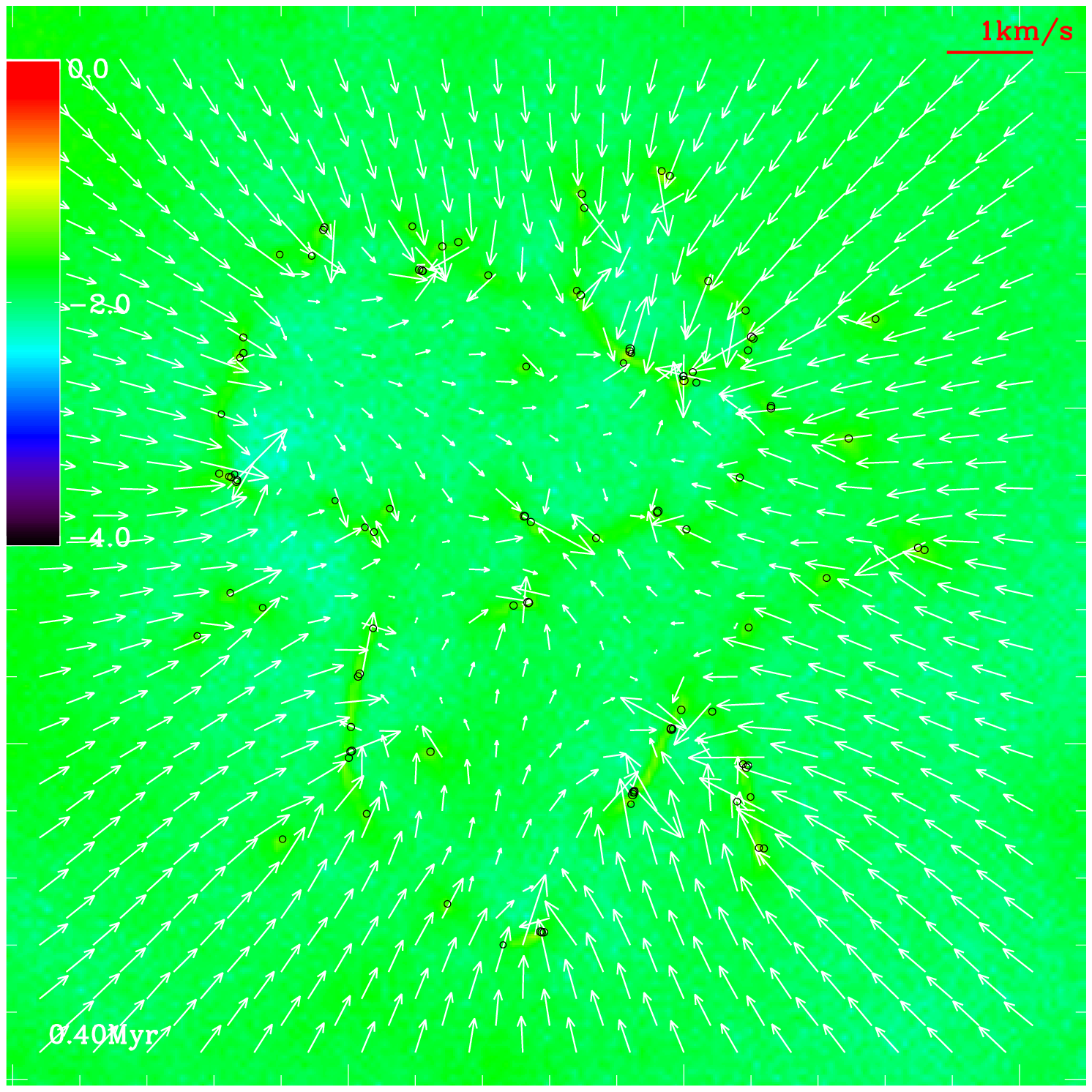}\\
	\includegraphics[scale=0.38]{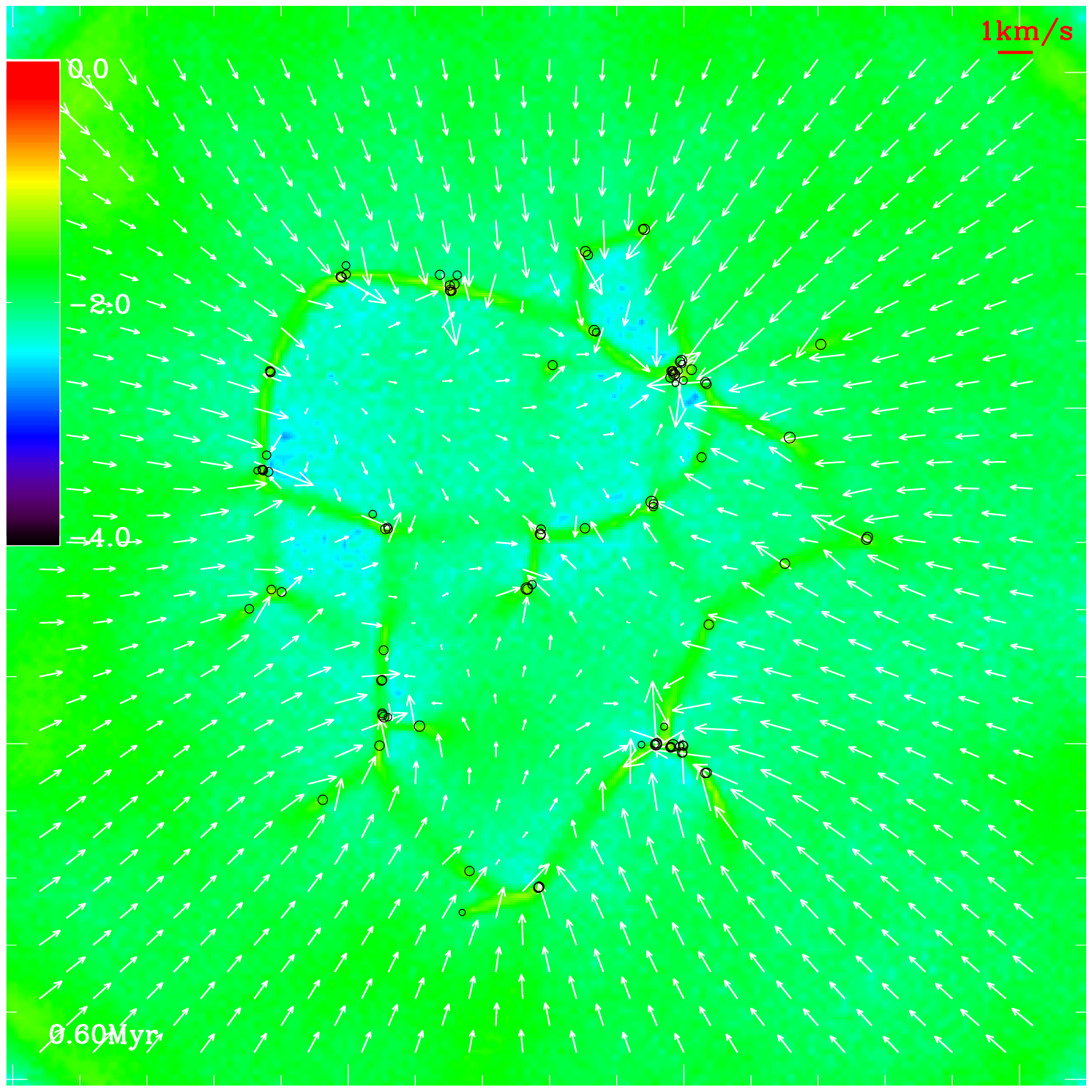}
	\includegraphics[scale=0.38]{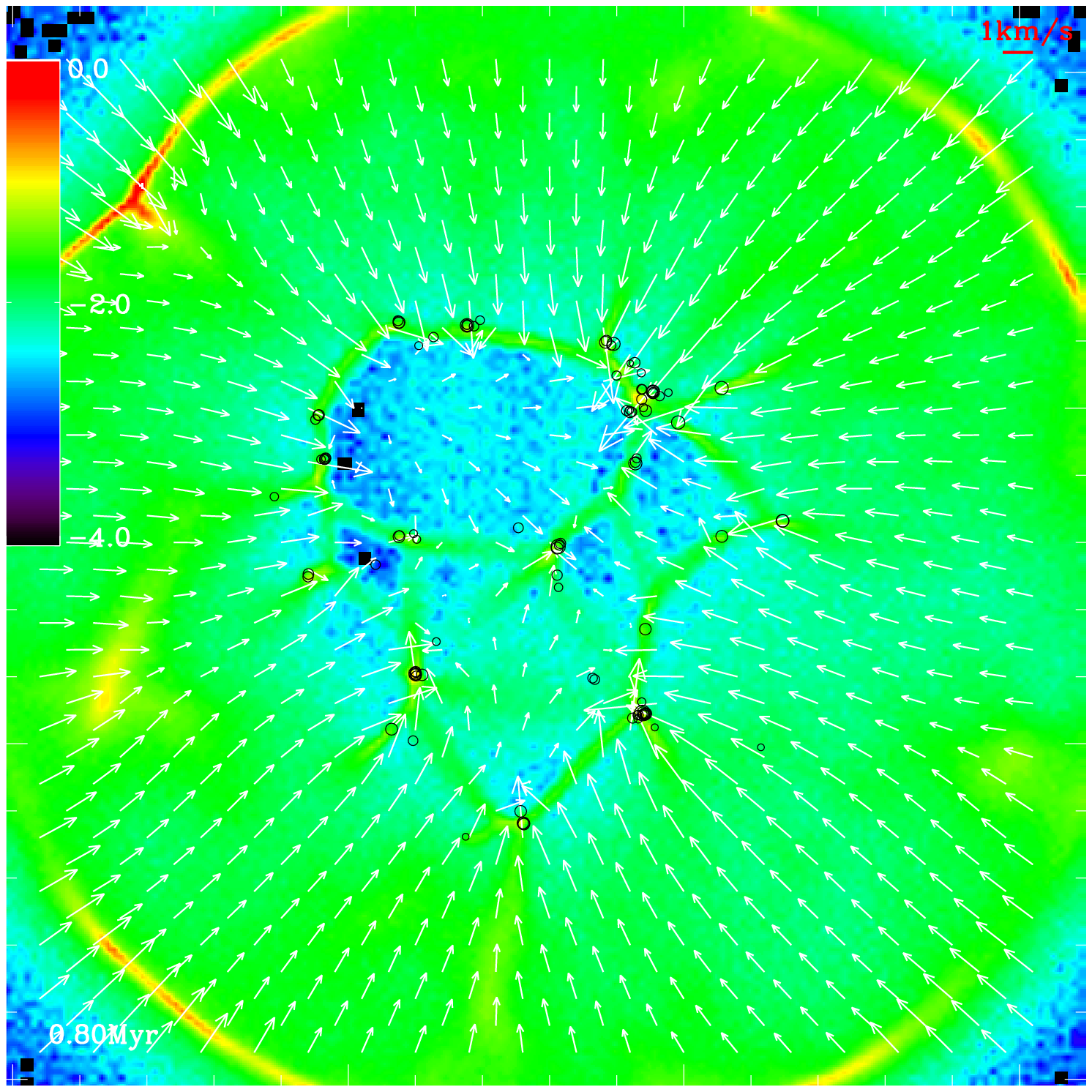}\\
	\includegraphics[scale=0.38]{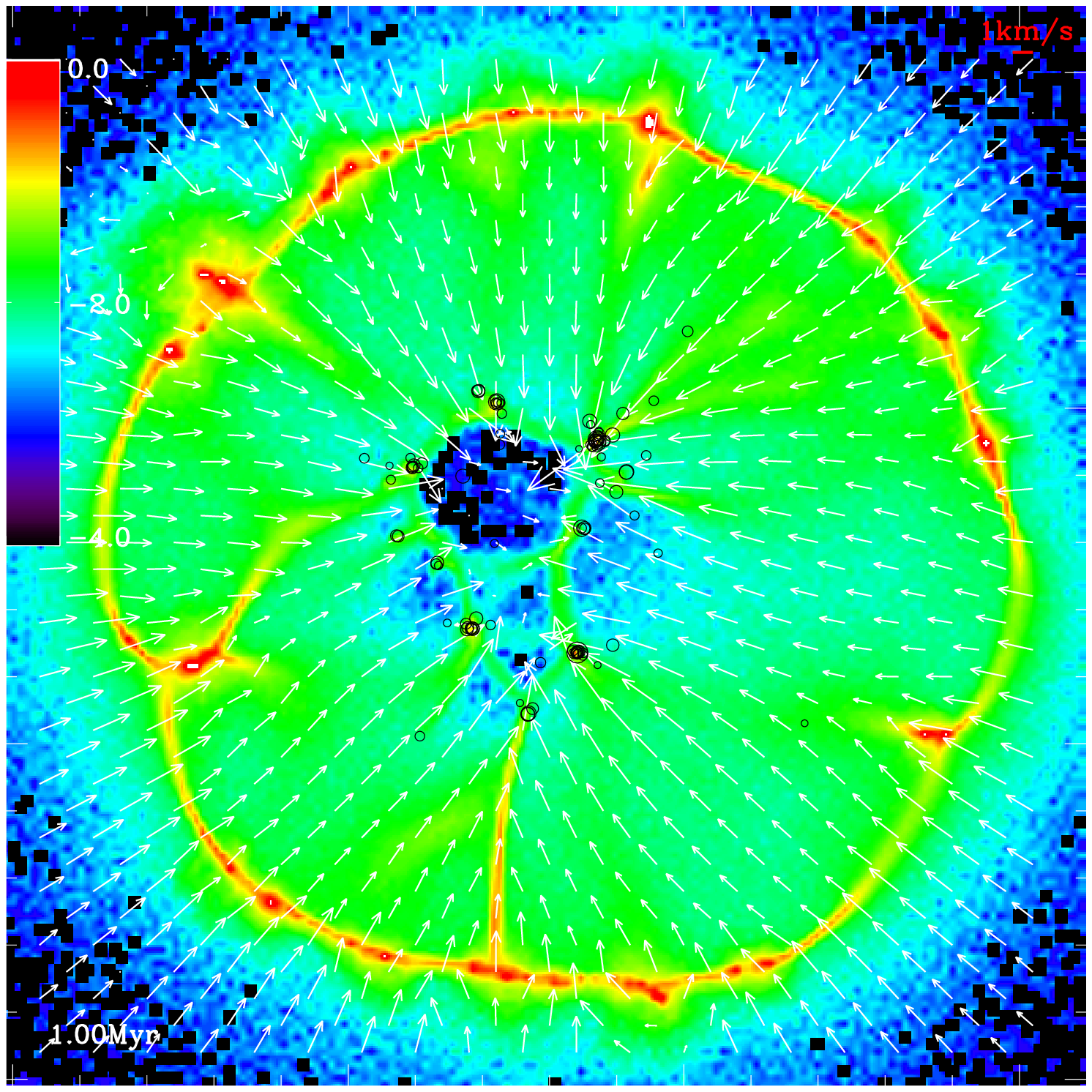}
	\includegraphics[scale=0.38]{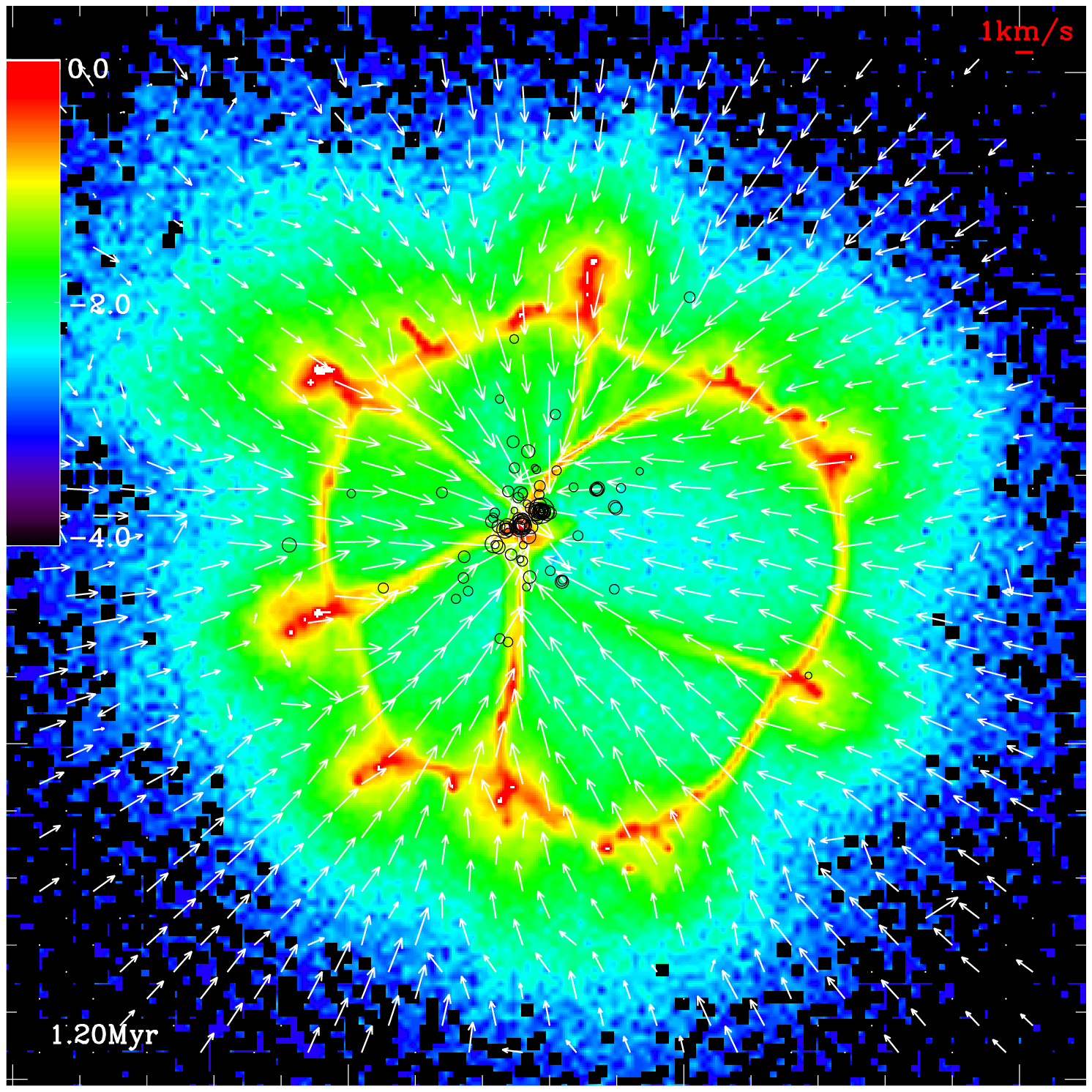}
	\caption{The top view of the inner part (each box is 2.4 by 2.4 pc) of the case with equal clump mass and background density fluctuation. The arrows indicate the velocity vectors of the gas, with 1$\kms$ marked on the upper right corner of each panel.  The colors correspond to the log of column density in $\rm{g~cm}^{-2}$.}
	\label{map_fluctuation}
	\end{center}
\end{figure}

\begin{figure}
	\includegraphics{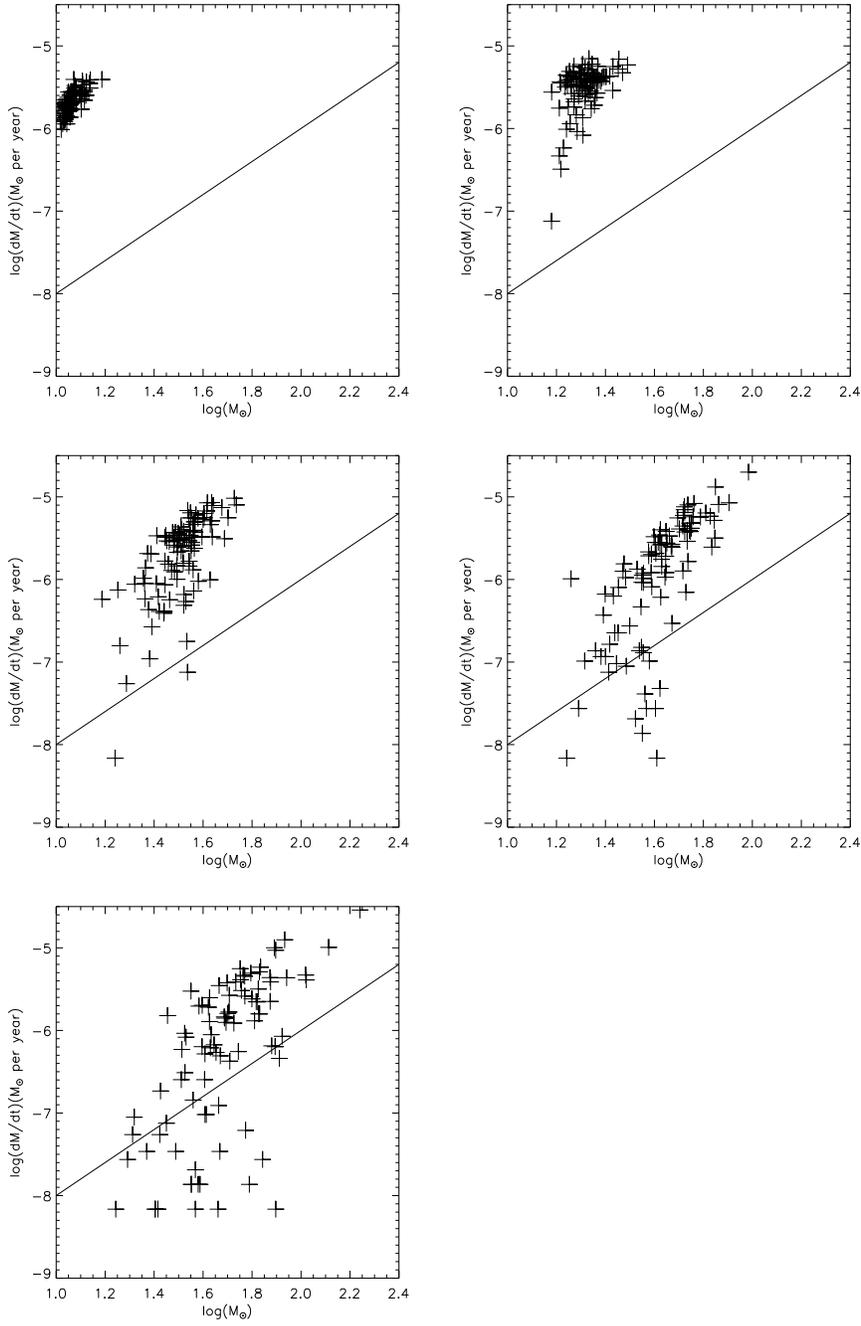}
	\caption{The accretion rate vs. mass of sinks at 0.2, 0.4, 0.6, 0.8 and 1.0 Myr in one of the runs of the fluctuating background case. The accretion rates for the higher-mass sinks follow $\dot{M} \propto M^2$; the surface density fluctuations have little effect (see Figure 4 for comparison).}
		\label{fluctuation_m2}
\end{figure}

\begin{figure}
	\includegraphics[scale=0.7]{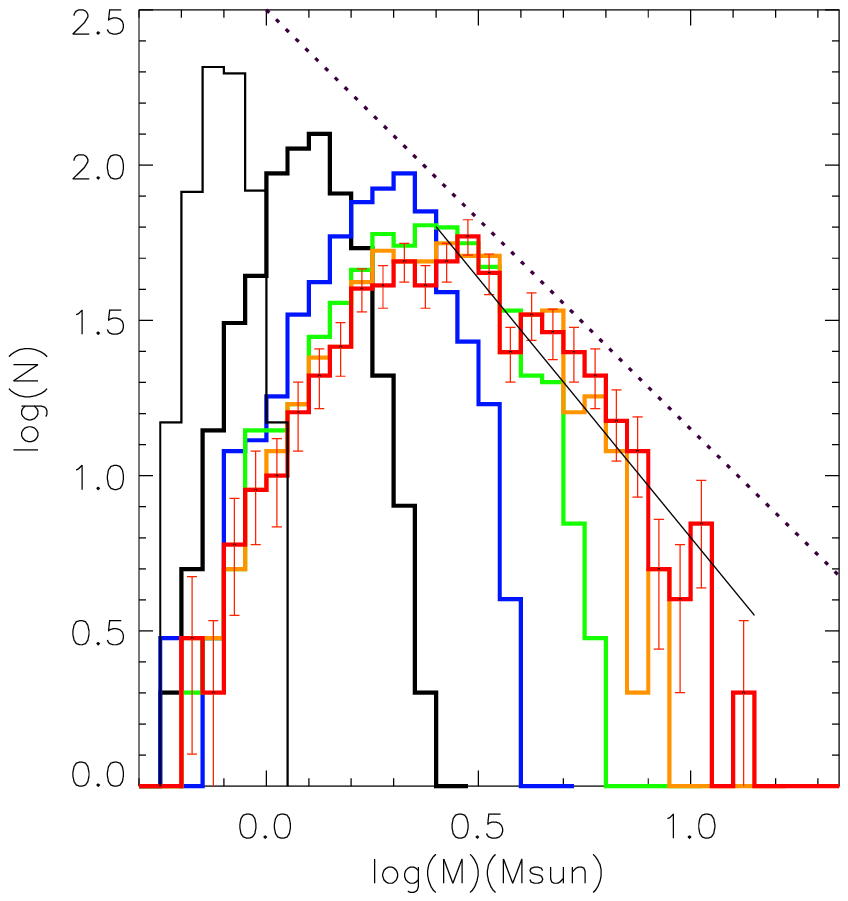}
	\includegraphics[scale=0.7]{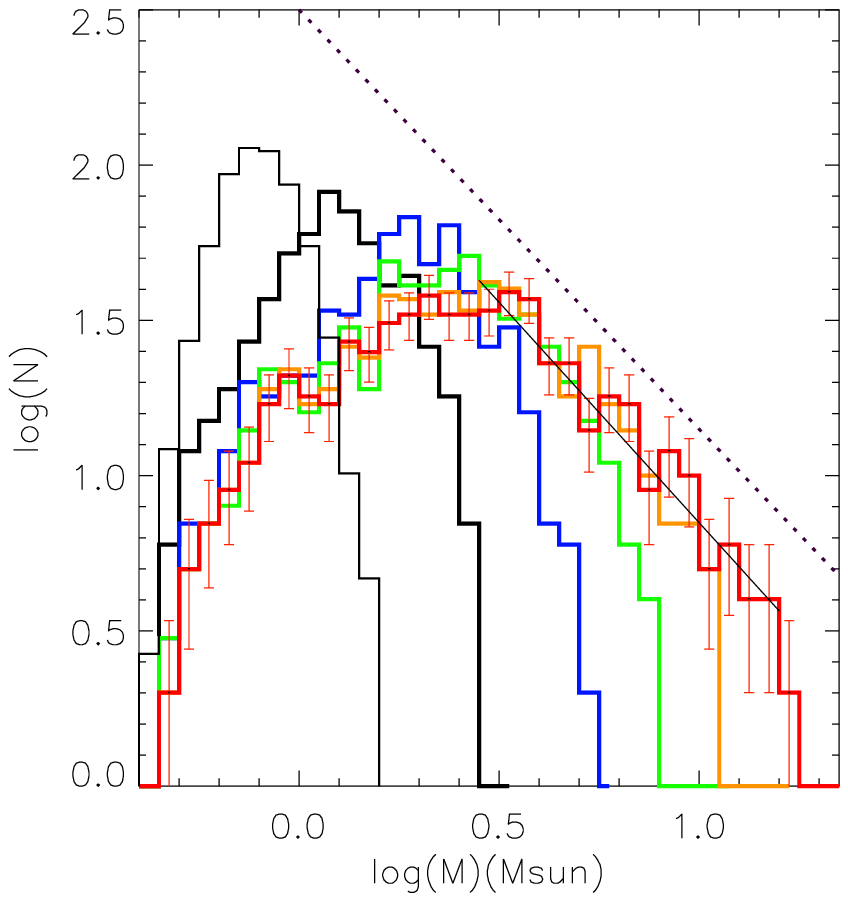}
	\includegraphics[scale=0.7]{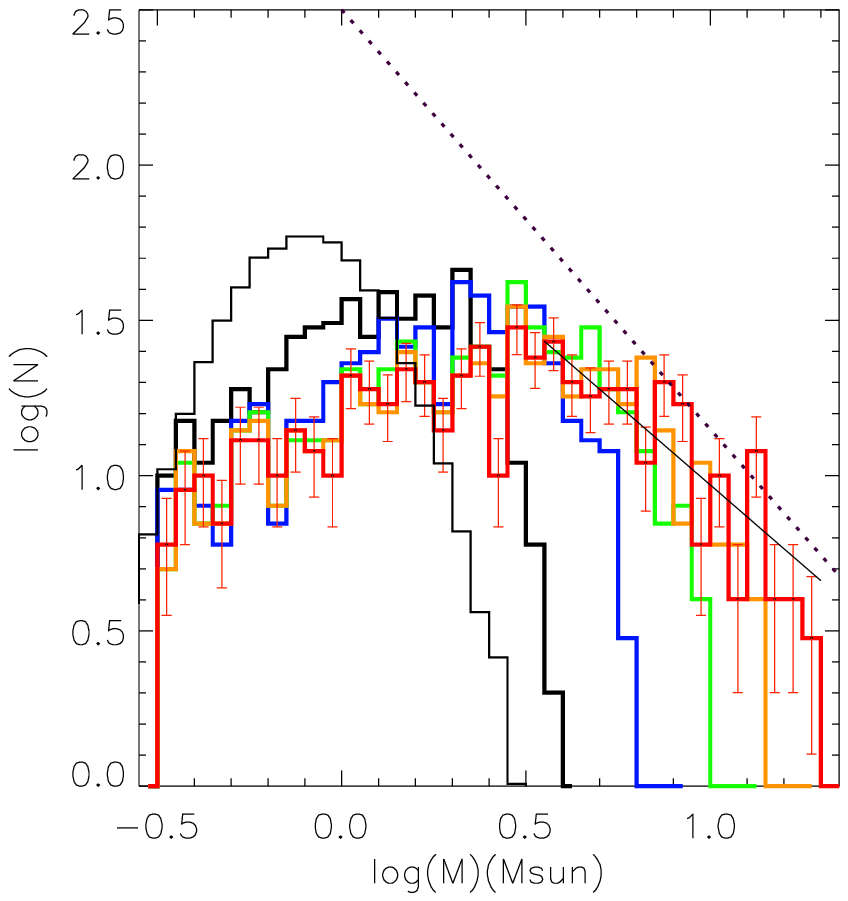}
	\caption{Resulting mass distributions for initial clump masses with Gaussian distributions (Top left: $\sigma = 0.05$; top right: $\sigma =0.1$; bottom left: $\sigma= 0.2$) with constant background density. The linear fit slopes are $-\Gamma$ = -1.66, -1.42 and -1.03, respectively. The thin solid lines  represent the initial clump mass distribution. The dotted lines represent the Salpeter slope $-\Gamma$ = -1.35.}
		\label{gauss}
\end{figure}

\begin{figure}
	\begin{center}	
\includegraphics[scale=0.7]{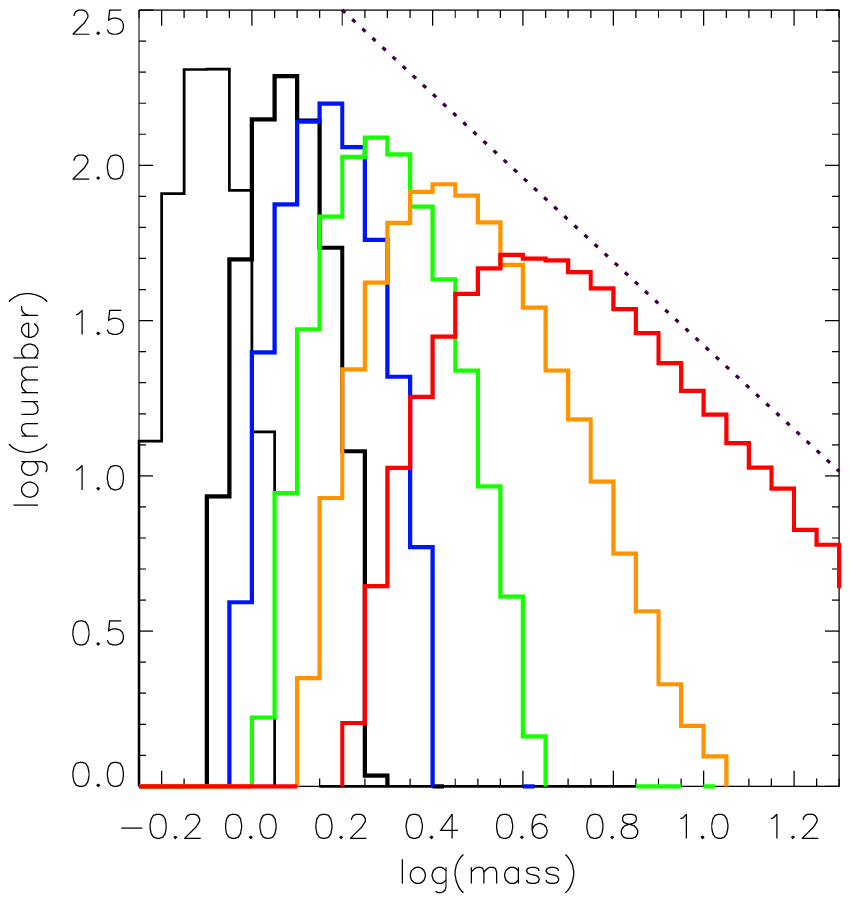}\\
\includegraphics[scale=0.7]{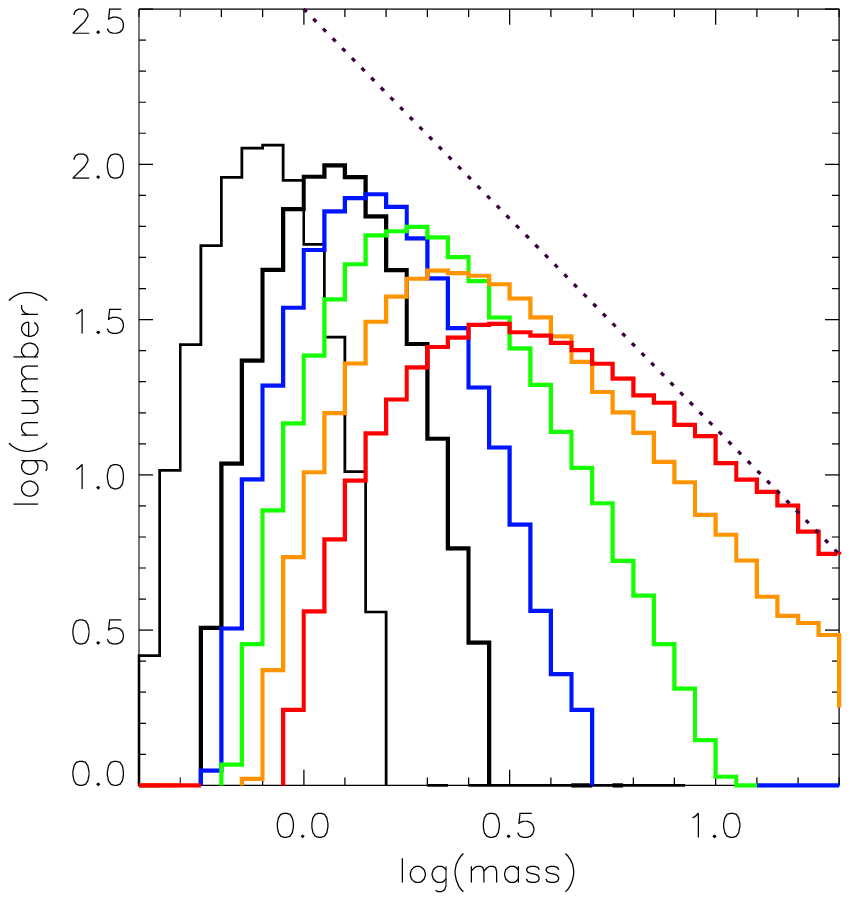} 

\caption{The mass distribution at different time for $\dot{M} = \alpha M^2$. The top figure has a narrower initial mass range ($\sigma$ = 0.05 dex) than the bottom figure($\sigma$ = 0.1 dex). The thin solid line represents the initial distribution of masses, and the colored lines represent the mass distribution in increments of $0.16 t_\infty= (\alpha M_0)^{-1}$. The dotted line represent the Salpeter slope $-\Gamma$ =  -1.35.}
	\label{Zinnecker}
	\end{center}
\end{figure}

\begin{figure}
	\includegraphics{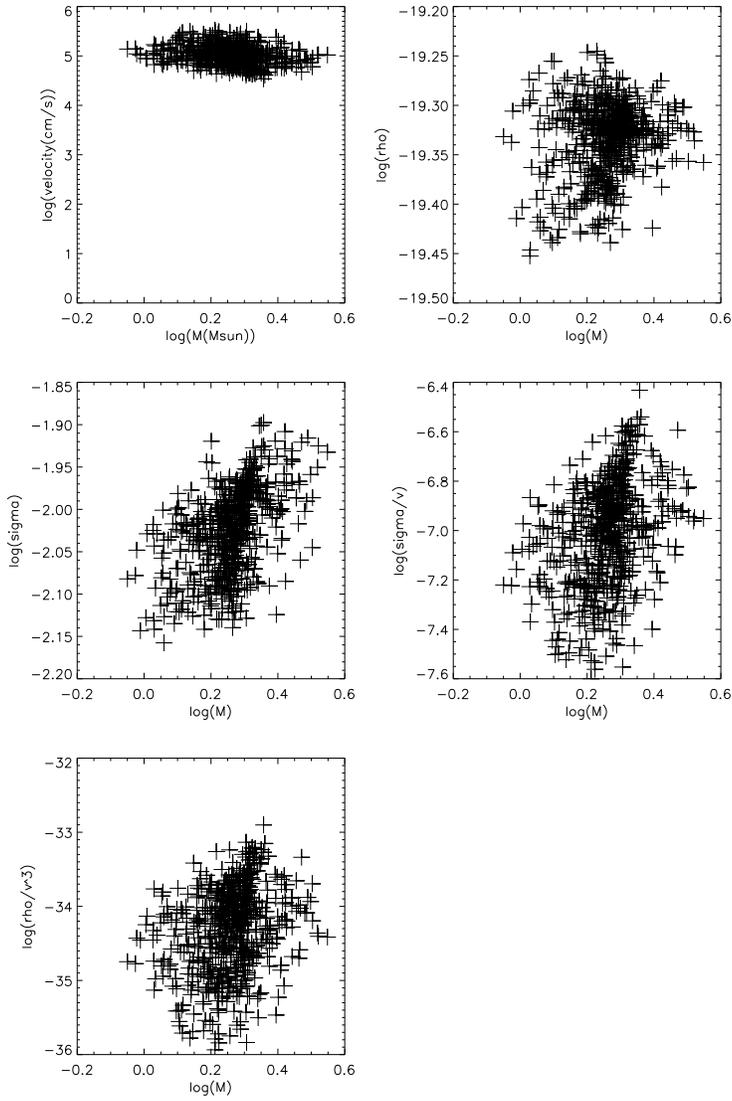}
	\caption{Mass of sinks vs. the gas properties around the sinks for all the cases in the equal initial mass, uniform density case at t = 0.6Myr. The gas properties are evaluated at $R_{acc}= 2GM/c_s^2$ from each sink. Top left: mass vs. gas velocity relative to the sink, top right: mass vs. gas density, middle left: mass vs. surface density, middle right: mass vs. surface density/velocity, bottom left: mass vs. $\rho/v^3$. There is no obvious correlation between the gas properties and the sink mass.}
		\label{correlation}
\end{figure}

\begin{figure}
	\includegraphics{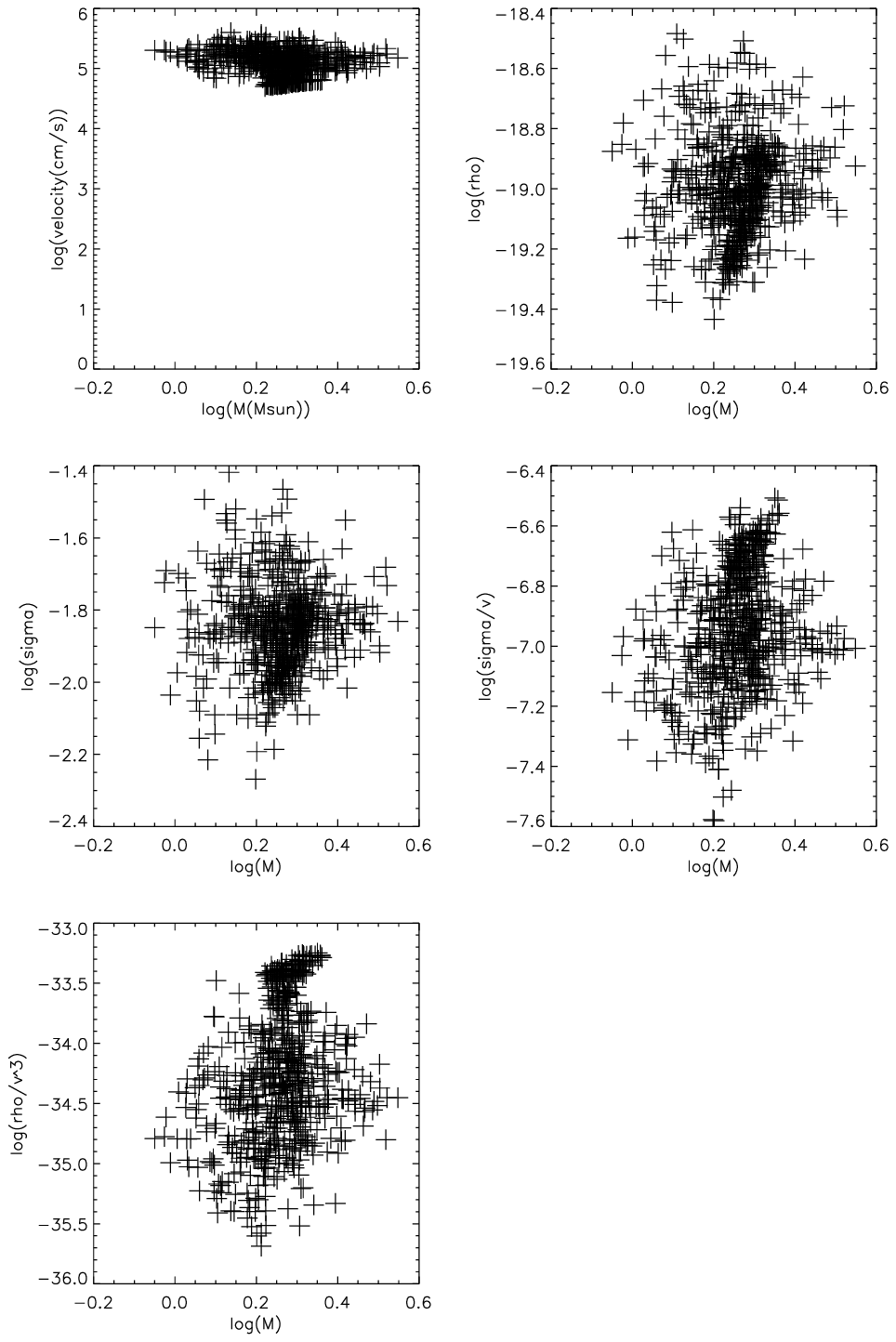}
	\caption{Mass of sinks vs. the gas properties around the sinks for all the cases in the equal initial mass, uniform density case at t = 0.6Myr. The gas properties are evaluated at $R= c_s \times 0.1Myr = 0.024$ pc from each sink. Top left: mass vs. gas velocity relative to the sink, top right: mass vs. gas density, middle left: mass vs. surface density, middle right: mass vs. surface density/velocity, bottom left: mass vs. $\rho/v^3$. There is no obvious correlation between the gas properties and the sink mass.}
		\label{correlation_024}
\end{figure}